\pdfoutput=1
\documentclass[twocolumn,superscriptaddress,aps,preprintnumbers,amsmath,amssymb,prd,nofootinbib]{revtex4-2}

\usepackage{amsmath}
\usepackage{amssymb}
\usepackage{amsfonts}
\usepackage{graphicx}
\usepackage{xcolor}
\usepackage{xfrac}
\usepackage{comment}
\usepackage{pifont}
\usepackage{physics}
\usepackage{fourier}
\usepackage{hyperref}
\usepackage{bm}
\usepackage{enumitem}
\usepackage{booktabs}

\definecolor{rossoferrari}{HTML}{D9073D}
\definecolor{mediumblue}{HTML}{0000CD}
\definecolor{forestgreen}{HTML}{228B22}
\definecolor{desy_blue}{HTML}{009EE2}
\definecolor{desy_orange}{HTML}{FD8800}
\definecolor{light_pink}{rgb}{1,0.4,0.4}
\definecolor{light_blue}{rgb}{0.284602,0.317763,0.963947}
\hypersetup{setpagesize=false,bookmarksnumbered=true,bookmarksopen=true, colorlinks=true,linkcolor=light_blue,urlcolor=rossoferrari,citecolor=rossoferrari,linktocpage=false}

\usepackage{slashed}

\renewcommand{\thefootnote}{\fnsymbol{footnote}}

\newcommand{\bea}{\begin{array}}
\newcommand{\eea}{\end{array}}
\newcommand{\beq}{\begin{eqnarray}}
\newcommand{\eeq}{\end{eqnarray}}

\newcommand{\lmk}{\left(}  
\newcommand{\rmk}{\right)}

\def\eq#1{Eq.~(\ref{#1})}

\newcommand{\SU}{\mathrm{SU}}
\newcommand{\PSU}{\mathrm{PSU}}
\newcommand{\ZN}{\mathbb{Z}_N}

\definecolor{orange}{RGB}{255,100,0}
\definecolor{rosepink}{RGB}{248,100,100}

\begin{document}

\title{Formation and scaling of $\mathbb{Z}_N$ strings for global $\mathrm{SU}(N)/\mathbb{Z}_N$ symmetry}

\author{Masaki Yamada}
\email{m.yamada@tohoku.ac.jp}
\affiliation{Department of Physics, Tohoku University, Sendai, Miyagi 980-8578, Japan}

\preprint{TU-1312}

\date{\today}

%%%%%%%%%%%%%%%%%%%%%%%%%%%%%%%%%%%%%%%%%%%%%%%%%%%%%%%%%%%%%%%%%%%%%%%%%%%%%%%%%%%%%%%%%%%%%%%%%%%%

\begin{abstract}
We numerically investigate networks of global $\mathbb{Z}_N$ strings with multi-string junctions in a scalar field model whose vacuum manifold is $\mathrm{PSU}(N)=\mathrm{SU}(N)/\mathbb{Z}_N$.
The construction of the model is motivated by the Higgs vacua of mass-deformed $\mathcal{N}=4$ supersymmetric Yang--Mills theory, commonly known as $\mathcal{N}=1^*$ theory, and provides a tractable effective description of string networks with baryon-vertex-like junctions.
We perform classical lattice simulations of the formation and evolution of these networks in a radiation-dominated universe.
For $N=2,3,$ and $4$, the time histories suggest that the total winding-length density approaches approximate scaling over the simulated interval, with no evidence for frustration.
For $N=5$ and $8$, we report densities at selected times.
For the parameters studied, the density normalization grows approximately in proportion to $N^2-1$, while non-minimal-charge components remain subdominant.
\end{abstract}

\maketitle

\renewcommand{\thefootnote}{\arabic{footnote}}
\setcounter{footnote}{0}

%%%%%%%%%%%%%%%%%%%%%%%%%%%%%%%%%%%%%%%%%%%%%%%%%%%
\section{Introduction}
%%%%%%%%%%%%%%%%%%%%%%%%%%%%%%%%%%%%%%%%%%%%%%%%%%%

Cosmic strings are one-dimensional topological defects that may have formed during phase transitions in the early Universe~\cite{Kibble:1976sj,Vilenkin:2000jqa}.
Once produced, a cosmic-string network evolves over cosmological time scales toward a scaling regime~\cite{Bennett:1987vf,Allen:1990tv} and can generate a stochastic gravitational-wave background through the dynamics of string loops~\cite{Vilenkin:1981bx,Vachaspati:1984gt}.
The recent evidence for a nanohertz stochastic gravitational-wave background reported by several pulsar timing array collaborations~\cite{NANOGrav:2023gor,EPTA:2023fyk,Reardon:2023gzh,Xu:2023wog,NANOGrav:2023hvm} has renewed interest in cosmic strings as a possible probe of high-energy physics beyond the Standard Model.

A particularly simple microscopic origin of cosmic strings is provided by a hidden pure Yang--Mills sector without fermions or Higgs fields~\cite{Yamada:2022imq,Yamada:2022aax,Yamada:2023thl}, as originally pointed out by Witten~\cite{Witten:1985fp}. In a pure Yang--Mills theory, color flux tubes appear in the confined phase as string-like excitations. If such a sector exists in the early Universe, the transition from the deconfined phase to the confined phase can lead to the formation of macroscopic color flux tubes over causally disconnected regions. These flux tubes then behave as cosmological strings. In the large-$N$ limit and in holographic descriptions, the reconnection probability of such strings can be suppressed, making their phenomenology similar to that of cosmic superstrings \cite{Polchinski:1988cn,Dvali:2003zj,Copeland:2003bj,Jackson:2004zg,Hanany:2005bc} 
and allowing them to fit the pulsar timing array signals~\cite{Yamada:2022imq,Yamada:2022aax,Ellis:2023tsl}. 
This mechanism does not require an elaborate brane-inflation scenario~\cite{Dvali:2003zj,Copeland:2003bj}, extra dimensions, or additional matter fields. It follows simply from the existence of a pure confining gauge theory. 
Depending on the gauge structure, color flux tubes have diverse
properties, including distinct center-charge sectors and string tensions~\cite{Douglas:1995nw,Hanany:1997hr,Witten:1998zw,Polchinski:2000uf,Klebanov:2000hb,Maldacena:2000yy,Vafa:2000wi} (see also Ref.~\cite{Yamada:2022imq}). 

\begin{figure}[t]
  \centering
  \includegraphics[width=0.92\linewidth]{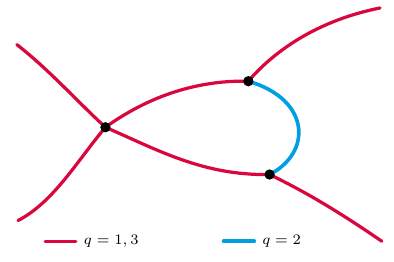}
  \caption{Schematic $N=4$ $\ZN$-string network.
  Red curves denote the unoriented unit-charge class $q=1,3$, while
  the blue curve denotes a $q=2$ string.  A four-string
  baryon-vertex-like junction can carry four unit strings, and the blue
  segment connects three-pronged center-neutral junctions.  Several
  strings are drawn as continuing outside the frame.}
  \label{fig:intro-schematic}
\end{figure}

The present paper is motivated by cosmic strings that may emerge in a pure
$SU(N)$ Yang--Mills theory.  The structure of confining strings in $SU(N)$ Yang--Mills
theory is richer than that of ordinary Abelian-Higgs strings.  The
fundamental confining strings carry a charge associated with the center
symmetry, and $N$ such strings can combine into a center-neutral
configuration.  Equivalently, $N$ strings can end on a localized
object called a baryon vertex~\cite{Witten:1998xy}
(see Fig.~\ref{fig:intro-schematic} for a schematic illustration).  For
$N=3$, this is closely analogous to the familiar baryonic
configuration in QCD, where three quarks are connected by color flux
tubes meeting at a Y-shaped junction.  
We are interested in the case without quarks, but the junction structure can still form. 
In the cosmological context, the existence of such junctions implies
that the string network is not simply a collection of independent
strings but is a network of $\ZN$-charged strings connected
by charge-conserving junctions.

This feature raises an important dynamical question.  Ordinary
cosmic-string networks are expected to approach a scaling regime, in
which the typical inter-string distance grows in proportion to the
horizon size and the number of long strings per horizon volume remains
of order unity.  This scaling behavior is crucial for making robust
cosmological predictions.  In contrast, a network with
charge-conserving junctions may in principle become frustrated.  Since
several strings can pull on each junction, the junctions might form a
long-lived, highly connected structure, leaving many strings and
junctions inside each horizon volume.%
\footnote{
Frustrated networks are familiar from
studies of non-Abelian string networks~\cite{Spergel:1996ai,McGraw:1997nx,Avgoustidis:2007aa}, where 
there are many different strings that cannot pass through or reconnect with each other. 
In contrast, in our case, the string vertices can annihilate because they are protected only by topological charge. 
We emphasize that neither our strings considered in the main part of this paper nor the color flux tubes in pure Yang--Mills theory fall into the category of non-Abelian strings, even though the symmetry or gauge group is non-Abelian. 
}
There is, however, another possibility.  The junctions may efficiently
annihilate or rearrange as the network evolves.  The network can then
reach a scaling regime, although with a normalization that differs from
the conventional string network.  Distinguishing between these two or
other unexpected possibilities is essential for predicting
observational signals.  In particular, the number of long strings per
horizon volume controls the string energy density and the efficiency of
loop production or radiation.  
Related networks with junctions have been studied in Refs.~\cite{Vachaspati:1986cc,Ng:2008mp,Copeland:2005cy,Hindmarsh:2006qn,Urrestilla:2007yw}, but it remains nontrivial whether similar scaling behavior occurs in the present model, especially for larger $N$.

The dynamics of such networks is difficult to determine from analytic arguments alone, because it depends on the nonlinear evolution of strings, vertices, and their interactions. A direct simulation of the confining transition in strongly coupled Yang--Mills theory is also beyond the scope of conventional real-time lattice methods. In this work, we instead take an effective field-theory approach. 
Motivated by the dual-superconductor picture of confinement~\cite{Mandelstam:1974pi,tHooft:1977nqb} and by its supersymmetric
realization through Seiberg--Witten electric--magnetic duality and
monopole condensation~\cite{Seiberg:1994rs,Seiberg:1994aj}
(see also, e.g., Refs.~\cite{Witten:1998zw,Polchinski:2000uf,Klebanov:2000hb,Maldacena:2000yy,Vafa:2000wi}), together with the intuition obtained from the $\mathcal{N}=1^*$ mass deformation of $\mathcal{N}=4$ supersymmetric Yang--Mills theory \cite{Donagi:1995cf,Dorey:1999sj,Polchinski:2000uf,Naculich:2001us}, 
we study a non-Abelian scalar model whose vacuum manifold is the projective special unitary group, 
\begin{equation}
  \PSU(N)=\SU(N)/\ZN.
\end{equation}
This vacuum structure admits $\ZN$-charged string solutions and
center-neutral junctions.  This model is not intended to reproduce all
microscopic properties of pure Yang--Mills theory.  Rather, it is
designed to isolate one key topological and dynamical ingredient
relevant for cosmology, such as the presence of $\ZN$ charge conservation and
baryon-vertex-like junctions in a string network.  In particular, the
strings are global, and the reconnection probability is inevitably
$\mathcal O(1)$ in our classical lattice simulations.

Field-theory simulations have long been used to study the scaling of
string networks~\cite{Vincent:1997cx,Yamaguchi:1998iv,Yamaguchi:1999yp}.
We study the real-time evolution of the global
$\ZN$-string networks by classical lattice simulations in an
expanding Universe.  In particular, we focus on $N=2,3,4,5,$ and $8$,
for which center-neutral junctions can involve two, three, four, five, and eight
unit-charge strings, respectively.  The case with $N=2$ is similar to ordinary cosmic strings without vertices. 
We perform three-dimensional
simulations on cubic lattices using parallel computation and reconstruct
the charge-resolved $\ZN$ string length.  This allows us to test
whether the network approaches a scaling solution or instead remains
frustrated by the junction structure.
For $N=2,3,$ and $4$, the time histories show no evidence for frustration over the simulated interval and suggest an approach toward approximate density scaling.
The normalization of the measured winding-length density increases approximately in proportion to $N^2-1$ for the parameters studied.

The rest of this paper is organized as follows. In Sec.~\ref{sec:model}, we introduce the effective scalar theory and discuss its vacuum structure, string charges, and baryon-vertex-like junctions. In Sec.~\ref{sec:string-profile}, we analyze the isolated-string ansatz and the radial structure of the non-Abelian core. In Sec.~\ref{sec:numerics}, we describe the lattice formulation, numerical implementation, and numerical results. 
Section~\ref{sec:discussion} is devoted to conclusions and discussion.

\section{Theoretical model}
\label{sec:model}

As discussed in the introduction, we consider a scalar field theory with
a global $\PSU(N)=\SU(N)/\ZN$ symmetry and investigate its transition
from a symmetric phase to the Higgs phase, in which the symmetry is
spontaneously broken.  The field content and scalar potential are
motivated by electric--magnetic duality and by Higgs-phase descriptions
of confinement, as specified below.

\subsection{Field content and scalar potential}

We denote by $T^a$ the Hermitian generators of $\mathfrak{su}(N)$ in the fundamental
representation, normalized as
\begin{align}
  [T^a,T^b]=i f^{abc}T^c,
  \\
  \Tr(T^aT^b)=\frac12\delta^{ab},
\end{align}
for $a,b,c=1,\ldots,N^2-1$.
The model contains three real adjoint scalar fields,
\begin{equation}
  \Phi_i(x)=\phi_i^a(x)T^a,
  \qquad i=1,2,3.
\end{equation}
Throughout this paper, the indices $i,j,k=1,2,3$ label the three
adjoint scalar fields, while $a,b,c,d=1,\ldots,N^2-1$ label adjoint
color components.  Greek indices denote spacetime directions, and
repeated flavor, adjoint, and spacetime indices are summed unless
otherwise stated.  The integer $N$ specifies the underlying $\SU(N)$
group, and $\ZN$ denotes its center.
For the values used in the simulations below, $N=2,3,4,5,$ and 8, the number of real field components is
\begin{equation}
  3(N^2-1)=9,\ 24,\ 45,\ 72,\ 189,
\end{equation}
respectively.

The Lagrangian density is
\begin{equation}
  \mathcal{L}
  = - \frac12\sum_{i,a}\partial_\mu\phi_i^a\partial^\mu\phi_i^a
  -V(\phi),
\end{equation}
with
\begin{align}
V = V_F + V_r. 
  \label{eq:potential}
\end{align}
The first term is the analogue of the $F$-term condition in the
$\mathcal{N}=1^*$ theory
\cite{Dorey:1999sj,Polchinski:2000uf,Naculich:2001us}, given by 
\begin{align}
  V_F (\phi)
  &=\sum_i \Tr(F_i^2)
  \\
  &=\frac{1}{2}\sum_{i,a}\left(F_i^a\right)^2,
\end{align}
where $F_i=F_i^aT^a$ and 
\begin{align}
  F_i^a
  &=m\phi_i^a
  -\frac{g}{2}\epsilon_{ijk}f^{abc}\phi_j^b\phi_k^c,
  \label{eq:Fterm}
\end{align}
where $m$ and $g$ are parameters.

Let $t_i$ be the generators of the irreducible $N$-dimensional representation of $\mathfrak{su}(2)$,
\begin{equation}
  [t_i,t_j]=i\epsilon_{ijk}t_k .
\end{equation}
A nonzero Higgs vacuum is
\begin{equation}
  \Phi_i^{({\rm vac})}=\frac{m}{g}t_i. 
  \label{eq:higgs-vacuum}
\end{equation}
Here and hereafter, we write $\Phi_i = \phi_i^a T^a$. 
The field-space radial norm is defined by
\begin{equation}
  \phi_r^2
  =\sum_{i,a} \left(\phi_i^a\right)^2
  =2\sum_i\Tr\left[\left(\Phi_i\right)^2\right], 
\end{equation}
and its Higgs-vacuum value is
\begin{equation}
  v^2
  =\left(\frac{m}{g}\right)^2\frac{N(N^2-1)}{2} .
  \label{eq:v-definition}
\end{equation}

The supersymmetric $F$-term potential also has the symmetric solution $\Phi_i=0$. 
For the numerical simulations of string formation, 
we add a radial stabilizing term that destabilizes the origin,
\begin{align}
V_r(\phi) = \frac{\lambda_r}{4}\left(\phi_r^2-v^2\right)^2.
\end{align}
We denote by $m_{\rm core}$ the tachyonic mass scale at the origin of the potential,
\begin{equation}
  m_{\rm core}^2\equiv \lambda_r v^2- m^2.
  \label{eq:m-core}
\end{equation}
We take $\lambda_r$ such that $m_{\rm core}^2 > 0$. 
The scalar field then rolls toward the Higgs vacuum even if it starts near the symmetric point.

In Appendix~\ref{sec:appA}, 
we calculate the curvature of the potential at the Higgs vacuum. 
The mass of the radial direction is
\begin{align}
  M_{\rm rad}^2= m^2+2\lambda_r v^2 .
  \label{eq:hessian-radial}
\end{align}
and the other massive directions have masses
\begin{align}
  M_{\ell-1}^2&=  m^2\ell^2,
  \qquad \ell=2,\ldots,N-1,
  \label{eq:hessian-spectrum-low}\\
  M_{\ell+1}^2&= m^2(\ell+1)^2,
  \qquad \ell=1,\ldots,N-1.
  \label{eq:hessian-spectrum-high}
\end{align}
Additionally, there are $N^2-1$ massless Goldstone directions.

For the parameters used in the simulations below, $m=1$ and $\lambda_r v^2=2$, so $M_{\rm rad}^2=5$ and the lightest massive Higgs-vacuum mode has $M_{\rm light}^2=4$, or $M_{\rm light}=2$.  This scale should be distinguished from $m_{\rm core}$ in Eq.~\eqref{eq:m-core}, which is defined from the curvature at the origin and controls the early roll-down.

\subsection{Vacuum manifold and global $\ZN$ strings}

The potential in Eq.~\eqref{eq:potential} admits a global $\SU(N)$ symmetry under which the scalar fields transform in the adjoint representation,
\begin{equation}
  \Phi_i(x)\mapsto U\Phi_i(x)U^{-1},
  \qquad U\in\SU(N).
\end{equation}
The center $\ZN\subset \SU(N)$ acts trivially on adjoint fields. Therefore the faithful global symmetry group acting on the scalar configuration is the projective special unitary group $\PSU(N)=\SU(N)/\ZN$. 
For the Higgs vacuum solution of \eq{eq:higgs-vacuum}, Schur's lemma implies that the only elements of $\SU(N)$ commuting with all $t_i$ are elements of the center. After quotienting by the center, the stabilizer in $\PSU(N)$ is trivial. Thus the vacuum manifold is $\PSU(N)$.
Since
\begin{equation}
  \pi_1(\PSU(N))=\ZN,
  \label{eq:pi1}
\end{equation}
cosmic strings can form after spontaneous symmetry breaking, and 
their string charge is classified by $q\in\ZN$
\cite{Vilenkin:2000jqa}.

A charge-$q$ string oriented along the $z$ direction can be described asymptotically in the transverse plane by
\begin{equation}
  \Phi_i(r,\varphi)\rightarrow h_q(\varphi)\Phi_i^{({\rm vac})}h_q(\varphi)^{-1}
  \qquad (r\rightarrow\infty),
  \label{eq:string-asymptotic}
\end{equation}
where $\varphi$ is the polar angle and $h_q(\varphi) \in \SU(N)$. 
Since the physical fields should be single-valued, 
$h_q(\varphi)$ must be single-valued as a $\PSU(N)$ map.  Its lift to
$\SU(N)$ may therefore close only up to a center element, such as
\begin{equation}
  h_q(2\pi)=\omega^q h_q(0),
  \qquad
  \omega=e^{2\pi i/N}.
  \label{eq:omega}
\end{equation}
This is the basic topological mechanism behind the global $\ZN$ strings simulated in this work.

The topological type of a string is its charge
\begin{equation}
  q\in\ZN .
\end{equation}
Orientation reversal maps
\begin{equation}
  q\mapsto -q \equiv N-q \pmod N .
\end{equation}
For example, for $N=3$ the two nontrivial charges $q=1,2$ are conjugate.  For $N=4$, the
charges $q=1$ and $q=3$ are conjugate, while $q=2$ is self-conjugate.  For $N=5$, the pairs
are $(1,4)$ and $(2,3)$.
We do not distinguish these conjugate pairs in our numerical simulations.

Charge conservation at a junction of $p$ strings requires
\begin{equation}
  \sum_{\alpha=1}^p q_\alpha=0\pmod N .
  \label{eq:junction-charge-conservation}
\end{equation}
In particular, $N$ unit-charge strings have total charge zero, as in
\begin{equation}
  \underbrace{1+\cdots+1}_{N\ {\rm times}}
  =
  N
  \equiv 0\pmod N ,
\end{equation}
and hence $N$ unit-charge strings can join at a vertex.

\section{Isolated string ansatz and radial profile}
\label{sec:string-profile}

We now discuss the structure of an isolated straight string in the
scalar-only $\PSU(N)=\SU(N)/\ZN$ model.  Although a closed-form analytic
solution is not available for the full non-Abelian profile, the
topological boundary conditions and the allowed core behavior can be
described explicitly.

\subsection{General condition for a string solution}

Consider a string oriented along the $z$ direction.  Let $(r,\varphi)$ be
polar coordinates on the transverse plane.  An example satisfying \eq{eq:omega} is generated by
\begin{equation}
  Y_q=
  \frac{1}{N}
  \mathrm{diag}
  (
  \underbrace{q,\ldots,q}_{N-q},
  \underbrace{q-N,\ldots,q-N}_{q}
  ),
  \qquad q=1,\ldots,N-1 .
  \label{eq:Yq}
\end{equation}
This matrix is traceless and satisfies
\begin{equation}
  \exp(2\pi iY_q)=\omega^q \mathbf 1_N,
  \qquad
  \omega=e^{2\pi i/N}.
\end{equation}
Thus
\begin{equation}
  h_q(\varphi)=\exp(i\varphi Y_q)
\end{equation}
is single-valued in $\PSU(N)$ but not in $\SU(N)$.  The asymptotic
boundary condition for a charge-$q$ string is \eq{eq:string-asymptotic}.

A general cylindrically symmetric ansatz can be written as
\begin{equation}
  \Phi_i(r,\varphi)
  =
  h_q(\varphi)\Psi_i(r)h_q(\varphi)^{-1},
  \label{eq:nonabelian-string-ansatz}
\end{equation}
where $\Psi_i(r)$ are radial matrix profiles. Defining the norm
\begin{equation}
  \|X\|^2=2\Tr(X^2),
\end{equation}
the energy per unit length becomes
\begin{equation}
  \mu_q
  =
  2\pi
  \int_0^{R_{\rm IR}} r\,dr
  \left[
  \frac12\sum_i\|\Psi_i'\|^2
  +
  \frac{1}{2r^2}\sum_i\|i[Y_q,\Psi_i]\|^2
  +
  V(\Psi)
  \right],
  \label{eq:mu-q-functional}
\end{equation}
where $R_{\rm IR}$ denotes the IR cutoff. 
The corresponding radial equations are
\begin{equation}
  \Psi_i''
  +
  \frac1r\Psi_i'
  -
  \frac{1}{r^2}[Y_q,[Y_q,\Psi_i]]
  =
  \frac{\partial V}{\partial \Psi_i}.
  \label{eq:matrix-profile-equation}
\end{equation}

The boundary condition at infinity is
\begin{equation}
  \Psi_i(\infty)=\Phi_i^{({\rm vac})} .
  \label{eq:string-bc-infinity}
\end{equation}
At the origin, finite energy requires that the physical field is
independent of the polar angle. Therefore
\begin{equation}
  [Y_q,\Psi_i(0)]=0 .
  \label{eq:core-commutant-condition}
\end{equation}
For the example of Eq.~\eqref{eq:Yq}, this condition means that
the allowed origin matrices are block diagonal with respect to the two
eigenspaces of $Y_q$, of dimensions $N-q$ and $q$, with total trace zero.
This condition is weaker than $\Psi_i(0)=0$.  Therefore, unlike in the
simplest Abelian string ansatz, the string core need not restore the
full symmetry in the present case.

\subsection{Numerical radial profile}

We determine the isolated-string profile by direct one-dimensional
relaxation of the functional in Eq.~\eqref{eq:mu-q-functional}.  As an example, we take $m=1$, $g=1$, and
$\lambda_r v^2=2$.  We expand the
three radial matrices $\Psi_i(r)$ in a Hermitian $\mathfrak{su}(N)$ basis,
discretize the interval $0\le r\le R_{\rm IR}$, and minimize the resulting
finite-difference energy.  At the origin, we vary only the matrix
components that satisfy Eq.~\eqref{eq:core-commutant-condition}, keeping
the angular-gradient term finite.  At the outer boundary, the matrices are
fixed to the Higgs vacuum.  The cases shown below use smooth interpolating
initial profiles.
We also checked that noisy or more strongly suppressed
initial profiles relax to the same origin value.

\begin{figure}[t]
  \centering
  \includegraphics[width=0.46\textwidth]{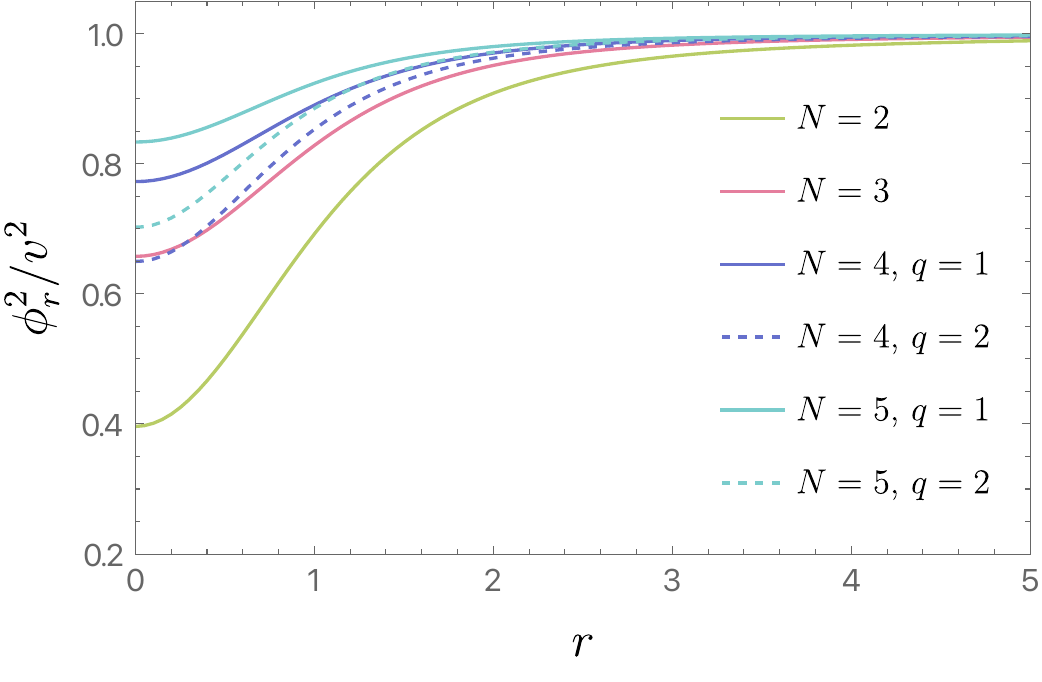}
  \caption{
  Numerical isolated-string radial profiles for the field-space radial norm
  $\phi_r^2/v^2$.  The horizontal axis is the cylindrical radius in units
  where $m=1$.  The figure labels the discrete charge by $q$.  The profiles show that the
  relaxed core does not, in general, drive $\phi_r^2$ to zero even at $r=0$.}
  \label{fig:radial-profile-phi-r}
\end{figure}

The relaxed profiles of the field-space radial norm are shown in
Fig.~\ref{fig:radial-profile-phi-r}.  The profiles approach the Higgs value
$\phi_r^2/v^2=1$ at large radius, but the origin value $\phi_r(0)$ remains nonzero. 
The resulting origin values are summarized in
Table~\ref{tab:radial-profile-core}.

\begin{table}[b]
  \centering
  \caption{Origin value and tension of the isolated-string radial
  profile for $m=1$, $g=1$, $\lambda_r v^2=2$, and
  $R_{\rm IR}=20$.  The last
  column lists the energy per unit length divided by $q(N-q)$,
  illustrating the approximate charge dependence expected from
  Eq.~\eqref{eq:log-tension-q}.}
  \label{tab:radial-profile-core}
  \begin{tabular}{cccc}
\toprule
    $N$ \  & $q$ \ & $\phi_r^2(0)/v^2$ \ & $\mu/[q(N-q)]$ \\
\midrule
    2 & 1 & 0.40 & 23.8 \\
    3 & 1 & 0.66 & 24.0 \\
    4 & 1 & 0.77 & 24.4 \\
    4 & 2 & 0.65 & 25.3 \\
    5 & 1 & 0.83 & 24.8 \\
    5 & 2 & 0.70 & 25.9 \\
\bottomrule
\end{tabular}
\end{table}

As in conventional global cosmic strings, the string core width is expected to be of the order of the inverse of the characteristic mass scale associated with the potential. In the parameter region of interest, it is typically set by the inverse core mass scale,
\begin{equation}
 \delta \sim \frac{1}{m_{\rm core}}. 
\end{equation}
The numerical examples in Fig.~\ref{fig:radial-profile-phi-r} are
consistent with this order-of-magnitude estimate.

The asymptotic energy density contains an unscreened angular-gradient
contribution.  For the example of Eq.~\eqref{eq:Yq}, one finds
\begin{equation}
  C_q
  =
  \sum_{i=1}^3
  \left\| i[Y_q,t_i]\right\|^2
  =
  2q(N-q).
  \label{eq:Cq}
\end{equation}
Thus the large-distance string tension contains a logarithmic global-string contribution from the second term in \eq{eq:mu-q-functional} of the form
\begin{equation}
  \mu_q
  \supset
  \pi
  \left(\frac{m}{g}\right)^2
  C_q
  \log\frac{R_{\rm IR}}{\delta}
  =
  2\pi
  \left(\frac{m}{g}\right)^2
  q(N-q)
  \log\frac{R_{\rm IR}}{\delta}.
  \label{eq:log-tension-q}
\end{equation}
Here
$\delta$ is the core width and $R_{\rm IR}$ is an infrared cutoff, such as
the horizon size or the mean string separation.  The equality
$C_q=C_{N-q}$ reflects the fact that $q$ and $N-q$ are string and
anti-string partners.
The numerical values in Table~\ref{tab:radial-profile-core} confirm that the ratio
$\mu/[q(N-q)]$ is nearly independent of $N$ and $q$ within the cases
shown.

\section{Classical lattice simulations}
\label{sec:numerics}

\subsection{Equations of motion in a radiation-dominated universe}
\label{subsec:eom-radiation}

We are interested in the formation and evolution of $\ZN$ strings in an expanding Universe.
We use conformal time $\eta$ and write the spatially flat Friedmann--Robertson--Walker metric as
\begin{equation}
  \dd s^2=a(\eta)^2\left(-\dd\eta^2+\dd\bm{x}^2\right),
\end{equation}
where $a(\eta)$ is the scale factor. 
We consider a radiation-dominated universe, for which
\begin{equation}
  a(\eta)=a_\ast\frac{\eta}{\eta_\ast},
  \qquad
  \mathcal{H}\equiv\frac{1}{a} \frac{\dd a}{\dd \eta}=\frac{1}{\eta},
  \label{eq:radiation-scale-factor}
\end{equation}
for a fixed $\eta_*$ with $a(\eta_*) = a_*$.
Throughout this work, $t_{\rm phys}$ denotes the cosmic age, including the age at initialization,
\begin{align}
  t_{\rm phys}(\eta)
  =\int_{0}^{\eta}a(\eta')\,\dd\eta'
  =\frac{a_\ast\eta^2}{2\eta_\ast}
  =\frac{a(\eta)\eta}{2}.
  \label{eq:tphys}
\end{align}

In addition to the physical evolution, we implement a 
core-growth, or fat-string, option, following the strategy commonly used
to keep defect cores resolved in expanding-lattice simulations
\cite{Press:1989yh,Yamaguchi:1999yp,Fleury:2015aca,Hindmarsh:2017qff,Hindmarsh:2019csc}.
We introduce a parameter $s$ by
evolving the fields with
\begin{align}
  S_s=\int \dd\eta\,\dd^3x\,
  \left[
  \frac{a^2}{2} \sum_{i,a} \left( \frac{\dd \phi_i^a}{\dd \eta} \right)^2
  -\frac{a^2}{2} \sum_{i,a} \left(\nabla\phi_i^a\right)^2
  -a^{2+2s}V(\phi)
  \right],
  \label{eq:physical-action}
\end{align}
where $V(\phi)$ is given by \eq{eq:potential}. 
The equation of motion is then given by 
\begin{equation}
  \frac{\dd^2 \phi_i^a}{\dd \eta^2}+2\mathcal{H} \frac{\dd \phi_i^a}{\dd \eta}
  -\nabla^2\phi_i^a
  +a^{2s} \frac{\partial V}{\partial\phi_i^a}=0.
  \label{eq:eom-second-order-physical}
\end{equation}
In component form,
\begin{equation}
  \frac{\partial V}{\partial\phi_i^a}
  =\left[
  mF_i^a
  -g\epsilon_{i jk}f^{abc}F_j^b\phi_k^c
  \right]
  +\lambda_r\left(\phi_r^2-v^2\right)\phi_i^a .
  \label{eq:dVdphi}
\end{equation}
For visualization, we use the energy density associated with the $s$-dependent evolution, given by 
\begin{equation}
  \rho(\bm{x})
  = \frac{1}{2a^2}\sum_{i,a} \lmk \frac{\dd \phi_i^a}{\dd \eta} \rmk^2 
  + \frac{1}{2a^2}\sum_{i,a} |\nabla\phi_i^a|^2 
  + a^{2s-2} V(\phi).
\end{equation}
The choice
\begin{equation}
  s=1
\end{equation}
reproduces the physical equation of motion.

The comoving mass scale around the global minimum is proportional to $a^s m_{\rm core}$, and hence the comoving core width scales as
\begin{equation}
  \delta_{\rm com}(\eta)\sim \frac{1}{a(\eta)^s m_{\rm core}}.
  \label{eq:core-width-s}
\end{equation}
Thus $s=1$ gives a physical core with $\delta_{\rm com}\propto a^{-1}$, whereas $s=0$ gives a fixed comoving core.  The $s=0$ runs are useful for following the network for a longer conformal time without losing core resolution, but they modify the physical core size and the logarithmic hierarchy of the global string.  We can use them to test the robustness of the approach to scaling.

\subsection{Numerical method}
\label{subsec:numerical-method}

We use a three-dimensional cubic lattice with periodic boundary conditions.  The main runs use $N_L^3=1024^3$ lattice points, while the $N=8$ run uses $512^3$ lattice points.  The comoving box size is $L$, and the lattice spacing is
\begin{equation}
  \Delta x=\frac{L}{N_L}.
\end{equation}
The Laplacian in the equation of motion is evaluated with a fourth-order centered difference, whereas gradients are evaluated using the second-order centered difference.
The evolution variable is conformal time $\eta$. We use the fourth-order symplectic method for time evolution.

To simplify the calculations, we approximate the phase transition as an instantaneous quench rather than a gradual physical cooling process. The dynamics of this rapid transition are simulated by initializing the Higgs field near the symmetric point ($\phi \sim 0$) and computing its subsequent roll-down into the symmetry-broken Higgs phase.
In the runs presented below, each real component of the scalar field is drawn from an uncorrelated Gaussian distribution with variance set by the input fluctuation amplitude.
The dependence on the initial fluctuation amplitude is examined in Sec.~\ref{subsec:simulation-checks}.
After the quench, the fields roll toward the global minimum.  Different causally disconnected regions select different points on the $\PSU(N)$ vacuum manifold, and strings form when the resulting field configuration winds nontrivially around $\pi_1(\PSU(N))=\ZN$.

Numerical simulations should end before either core-resolution or finite-volume effects become important.  The window must satisfy
\begin{equation}
  \delta_{\rm com} (\eta) \gtrsim \Delta x,
  \qquad
  \eta \le L/2.
  \label{eq:scaling-window-cuts}
\end{equation}
The second condition avoids the regime in which massless radiation and the string network have crossed the periodic box.

If both physical-core and fixed-comoving-core runs are used, the former define the physical result while the latter serve as a long-time robustness check.  Since the fixed-comoving-core evolution keeps the core resolved for longer in lattice units, it can test whether the winding density, charge-resolved densities, and local scaling exponent continue toward a plateau after the time at which a physical-core run would become under-resolved.

\subsection{Scaling parameters}

To identify the string position and its length, 
we count the number of plaquettes pierced by strings. 
The detailed numerical algorithm is explained in Appendix~\ref{sec:appC}. 
We group conjugate charges ($q \leftrightarrow N-q$) into unoriented string types and 
denote by $n_q$ the number of pierced plaquettes with charge
$q=1,\ldots,N/2$ for even $N$ and $q=1,\ldots,(N-1)/2$ for odd $N$.

For an isotropic string network, the plaquette-piercing count estimates a
Manhattan length, i.e. a length measured along lattice axes.  We convert it
to an approximate Euclidean comoving length by including the isotropic
factor $2/3$,
\begin{equation}
  \ell_{{\rm com},q}=\frac{2}{3}\Delta x\, n_q.
  \label{eq:winding-length}
\end{equation}

We then define the dimensionless scaling parameter associated with the
string length in charge class $q$ by
\begin{equation}
  \zeta_q(t)=t_{\rm phys}^2\frac{a \ell_{{\rm com},q}}{a^3 L^3}.
  \label{eq:zeta-charge}
\end{equation}
The scaling parameter for the total length is defined as 
\begin{align}
	  \zeta(t)&=\sum_q \zeta_q (t),
	  \label{eq:zeta-winding}
\end{align}
where the summation is taken over the unoriented nontrivial charge classes
defined above.
This observable includes all detected winding length, including closed loops.
We also use the fractional length
\begin{equation}
  R_q
  =
  \frac{\zeta_{q}}{\zeta}. 
  \label{eq:charge-ratio}
\end{equation}

A string network is said to scale when the mean string separation grows in proportion to the horizon scale.
Because $a\propto\eta$ and $t_{\rm phys}\propto\eta^2$ in radiation domination, scaling implies that $\zeta(t)$ should approach a plateau.%
\footnote{A related distinction between the coarsening exponent and the density normalization has been discussed for overdamped nematic $\mathbb{Z}_2$ defect networks~\cite{Efstratiou:2026kcb}.}

\subsection{Simulation parameters and numerical checks}
\label{subsec:simulation-checks}

We perform simulations for $N=2,3,4,5,$ and 8.
We use units in which $m=1$ and take $g=1$.
The radial coupling is chosen so
that the curvature at the origin, $-m_{\rm core}^2$, 
is equal to $-m^2$.  Equivalently,
\begin{equation}
  \lambda_r=\frac{2 m^2}{v^2},
\end{equation}
which gives $\lambda_r=2/3,\, 1/6,\,1/15,\,1/30,$ and $1/126$ for
$N=2,3,4,5,$ and 8, respectively.

The numerical and initial-condition parameters given below are benchmark values used unless otherwise stated.
We take $L=150$ and $N_L=1024$ ($512$ for $N=8$).
The conformal-time step is $\Delta\eta=0.02$.
The initial conformal time $\eta_0$ sets the initial ratio between the horizon scale and the string-core width.
We take $\eta_0=10$, normalize the scale factor to $a(\eta_0)=a_0=1$, and set the initial fluctuation amplitude to $0.1$.

We assess the sensitivity to numerical settings using runs with $N=3$ and $s=0$.
Reducing $\Delta\eta$ from $0.02$ to $0.01$ changes $\zeta$ by less than $0.2\%$ at all common output times in $30\leq\eta\leq50$.
Varying $L$ over $100$, $150$, and $200$ at fixed $N_L=1024$ gives agreement within approximately $3.4\%$ at $\eta=50$.
This comparison probes finite-volume and spatial-discretization effects together, since $\Delta x=L/N_L$ also changes.
Increasing $N_L$ from $512$ to $1024$ at fixed $L=100$ with $\Delta \eta = 0.05 \to 0.02$ gives a difference of approximately $3.1\%$ at $\eta=50$.
These checks indicate sensitivity at the few-percent level or below to the numerical settings tested.

Changing the initial fluctuation amplitude or $\eta_0$ instead changes the physical initial conditions of the quench.
For $N=3$ and $s=0$, reducing the amplitude from $0.1$ to $0.01$ at fixed $L=100$ increases $\zeta$ by approximately $12.1\%$ at $\eta=50$.
A smaller amplitude places the field closer to the symmetric point and delays its roll toward the broken phase and the formation of strings.
The networks therefore have different formation histories and different amounts of time to relax by a given $\eta$.
For the same $N$ and $s$, increasing $\eta_0$ from $10$ to $20$ at fixed $L=150$ increases $\zeta$ by approximately $42.0\%$ at $\eta=70$.
Both increases are consistent with delayed formation and relaxation of the network because the comoving core width is fixed and the horizon-to-core ratio is the same in the two runs at a given $\eta$ for $s=0$.

We also estimate the core width from the core volume fraction for $N=3$, using $N_L=688$ and $L=100$. 
Let $f_{\rm core}$ be the volume fraction satisfying $\phi_r^2<c_r v^2$, with $c_r=0.8$.
Approximating this region as cylinders with total comoving length $\ell_{\rm com}$, we define the effective radius and diameter by
\begin{align}
  r_{\rm eff,com}&=\sqrt{\frac{f_{\rm core}L^3}{\pi\ell_{\rm com}}},
  &w_{\rm eff,com}&=2r_{\rm eff,com}.
  \label{eq:effective-core-width}
\end{align}
Here $\ell_{\rm com}$ is the total winding length.

Table~\ref{tab:core-resolution} compares the measured diameters with the $a^{-1}$ rescaling expected for $s=1$ relative to $s=0$.
The $s=1$ diameters are approximately $11\%$, $9\%$, and $5\%$ smaller than the rescaled $s=0$ values at the three times, respectively.
Thus the measurements are approximately consistent with the expected scale-factor dependence.
Even at $a\Delta x\,m\simeq0.73$, the threshold-defined core therefore retains a diameter of more than two lattice spacings.
At $\eta=50$, the main $1024^3$ simulations with $L=150$ have $a\Delta x\,m=5\times150/1024\simeq0.73$, almost identical to the value in the $688^3$ core measurements at the same time, so comparable core resolution is expected.
At $\eta=70$, the physical lattice spacing increases to $a\Delta x\,m=7\times150/1024\simeq1.03$.
Assuming that the physical effective core diameter $w_{\rm eff,phys}=a w_{\rm eff,com}$ remains approximately constant for $s=1$, the measured diameter at $\eta=50$ implies $w_{\rm eff,phys}/(a\Delta x)\simeq2.26(5\times100/688)/(7\times150/1024)\simeq1.6$ at $\eta=70$ in the main simulations.
The late-time $s=1$ results, particularly the $\eta=70$ entries in Table~\ref{tab:scaling}, are obtained at the core-resolution limit.
The spatial-resolution accuracy of the $s=1$ evolution has not been tested, and these width estimates do not constitute a convergence test.

\begin{table}[t]
\centering
\caption{Effective comoving core diameters in lattice units for $N=3$, $c_r=0.8$, $N_L=688$, and $L=100$.
The scale factor is $a=\eta/10$.
Writing $w_s=w_{\rm eff,com}^{(s)}$, the third column gives the $s=0$ diameter divided by $a$ for comparison with the measured $s=1$ diameter in the last column.}
\begin{tabular}{cccc}
\toprule
$\eta$ & $w_0/\Delta x$ & $w_0/(a\Delta x)$ & $w_1/\Delta x$ \\
\midrule
40 & 13.06 & 3.27 & 2.92 \\
45 & 12.54 & 2.79 & 2.55 \\
50 & 11.94 & 2.39 & 2.26 \\
\bottomrule
\end{tabular}
\label{tab:core-resolution}
\end{table}

\subsection{Numerical results}
\label{sec:output}

\begin{figure}[t]
    \centering
    \includegraphics[width=0.34\textwidth]{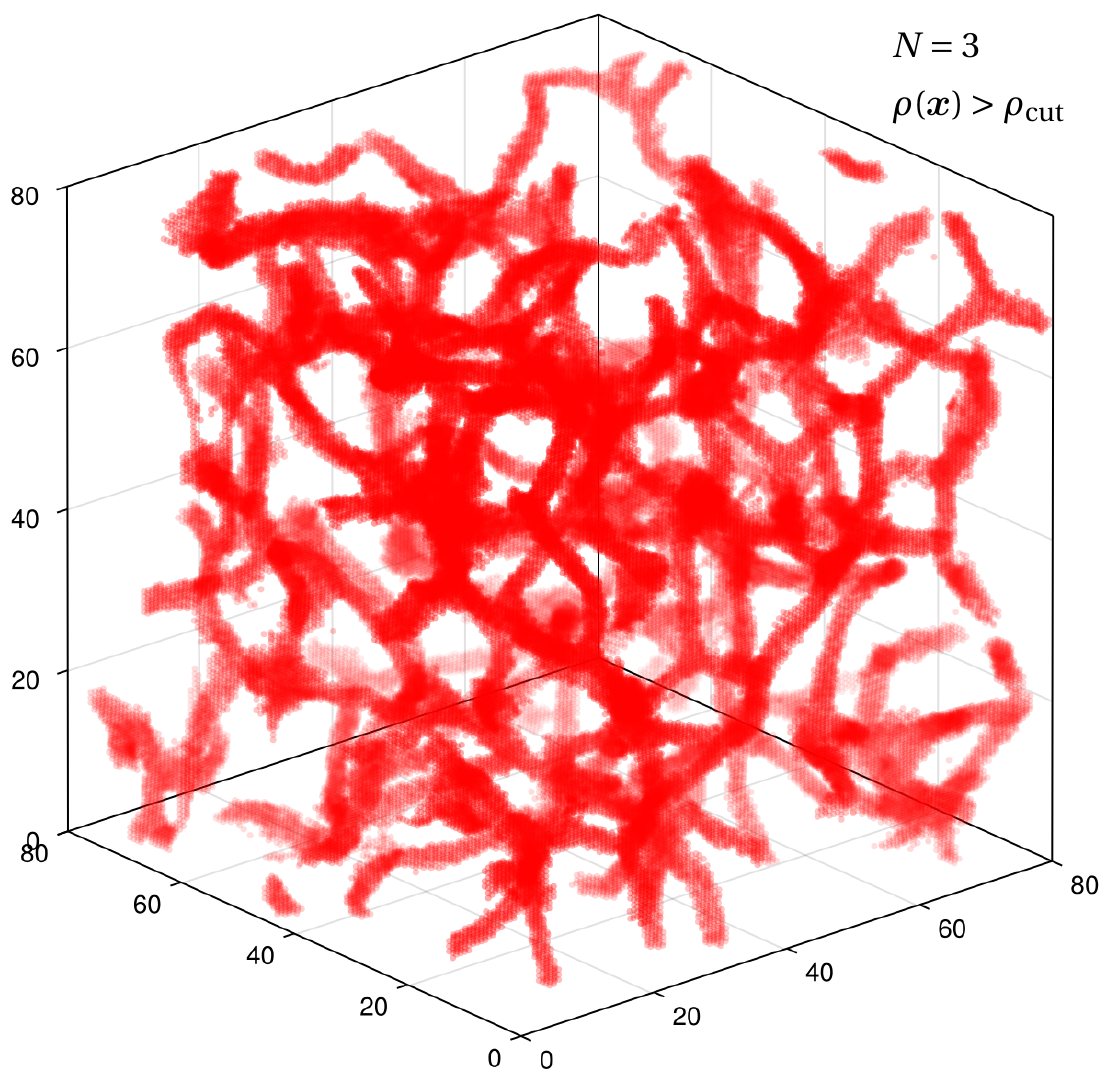}
    \\
    \includegraphics[width=0.34\textwidth]{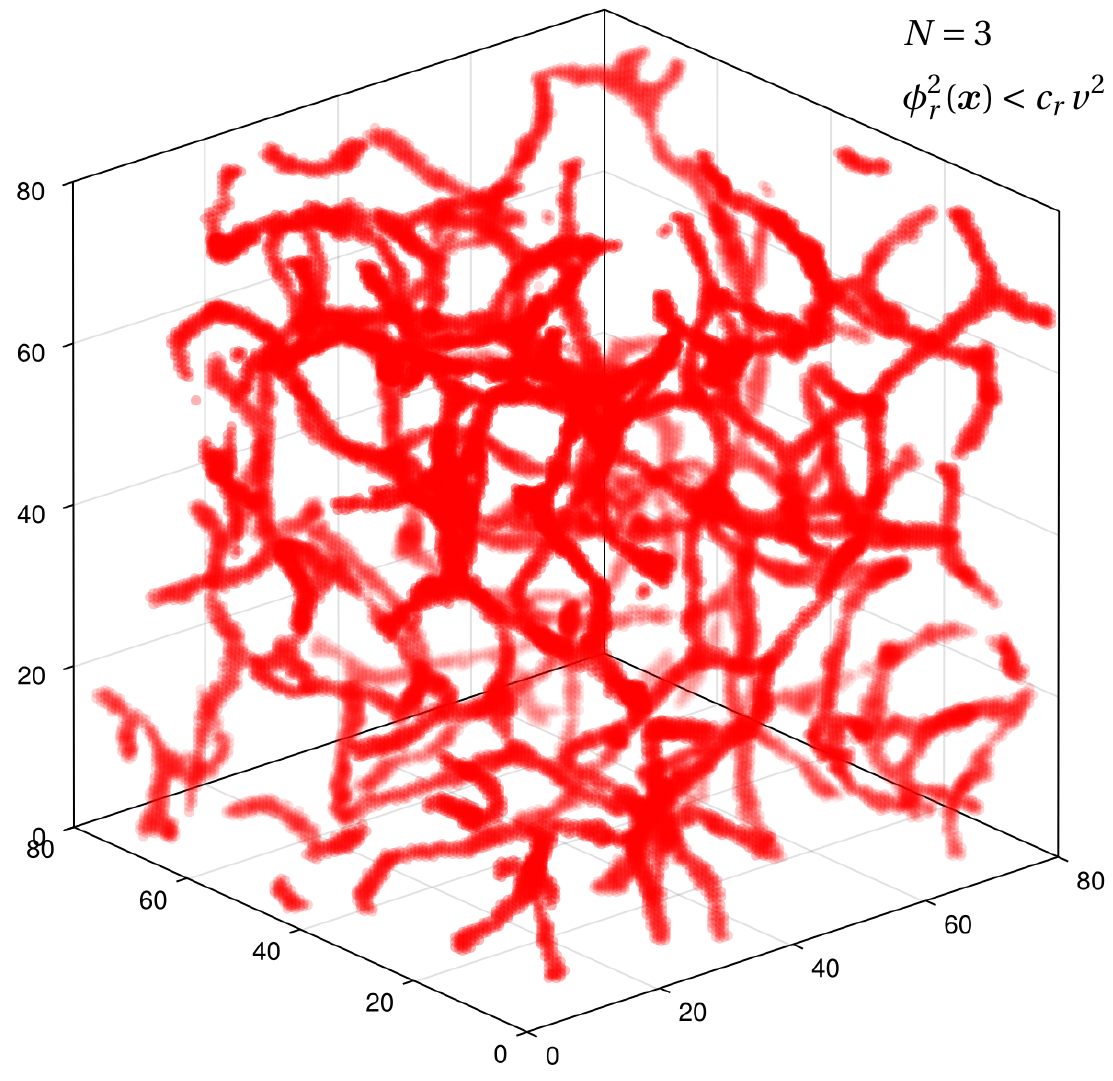}    
    \\
    \includegraphics[width=0.34\textwidth]{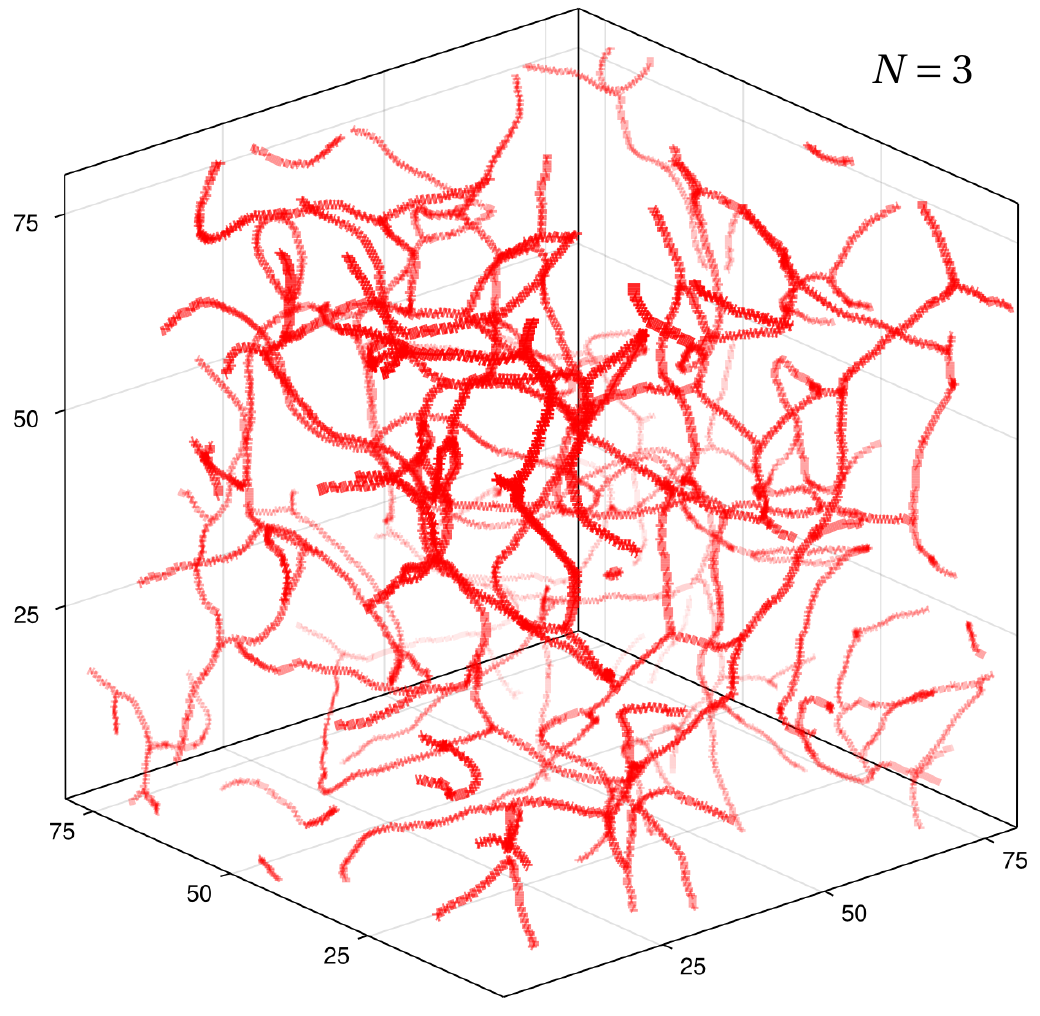}    
    \caption{
    Snapshot of an $N=3$ network in a smaller $256^3$ run with
    $L=80$ and $s=0$.  From top to bottom: high-energy regions
    $\rho>\rho_{\rm cut}$, radial-core regions
    $\phi_r^2<c_r v^2$, and the charge-resolved
    $\ZN$ string segments reconstructed from plaquette winding.  For
    $N=3$, the two nontrivial oriented charges are conjugate and are
    shown as a single red unoriented string type.}
    \label{fig:snapshot1}
\end{figure}

\begin{figure}[t]
    \centering
    \includegraphics[width=0.34\textwidth]{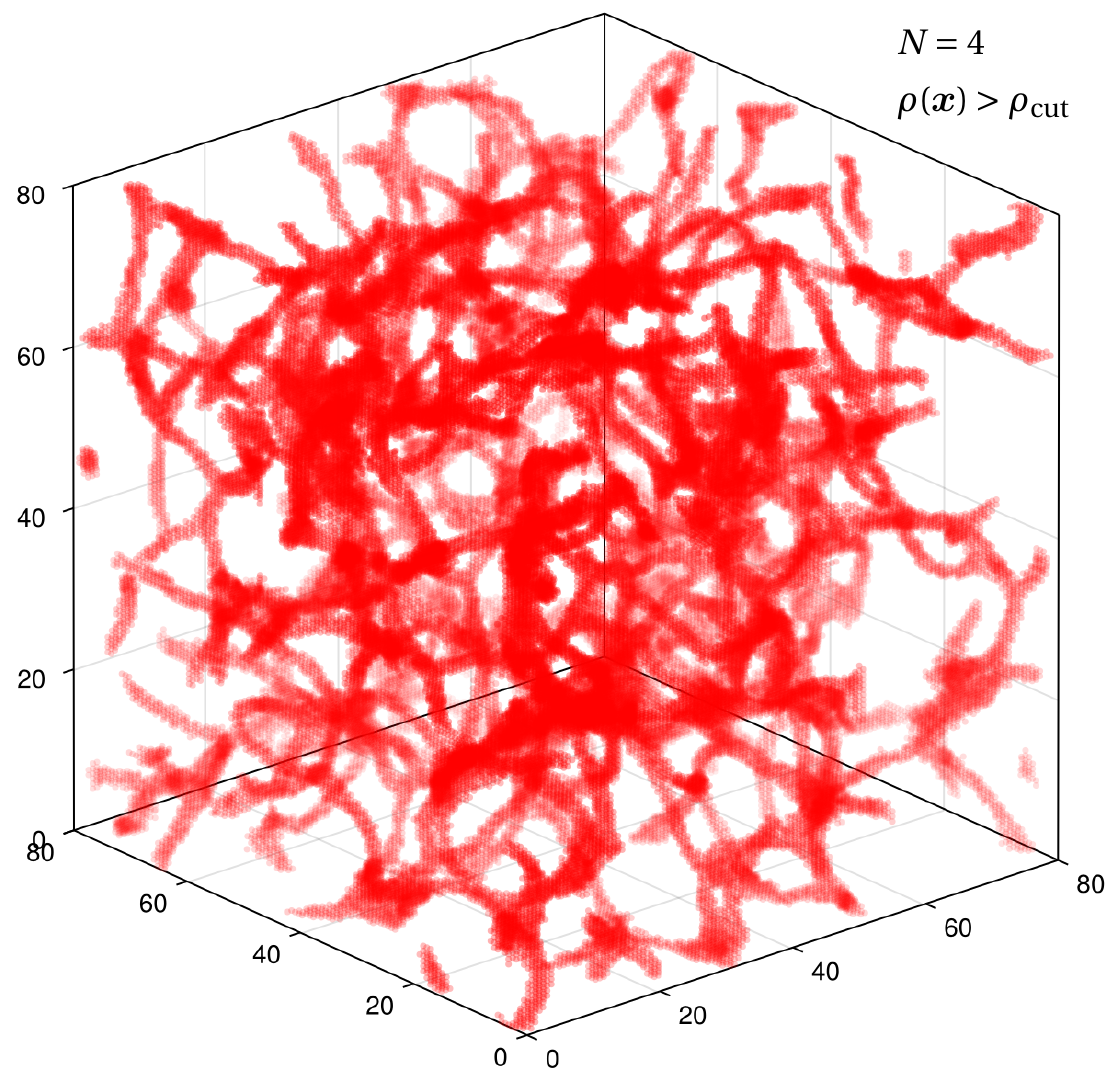}
    \\
    \includegraphics[width=0.34\textwidth]{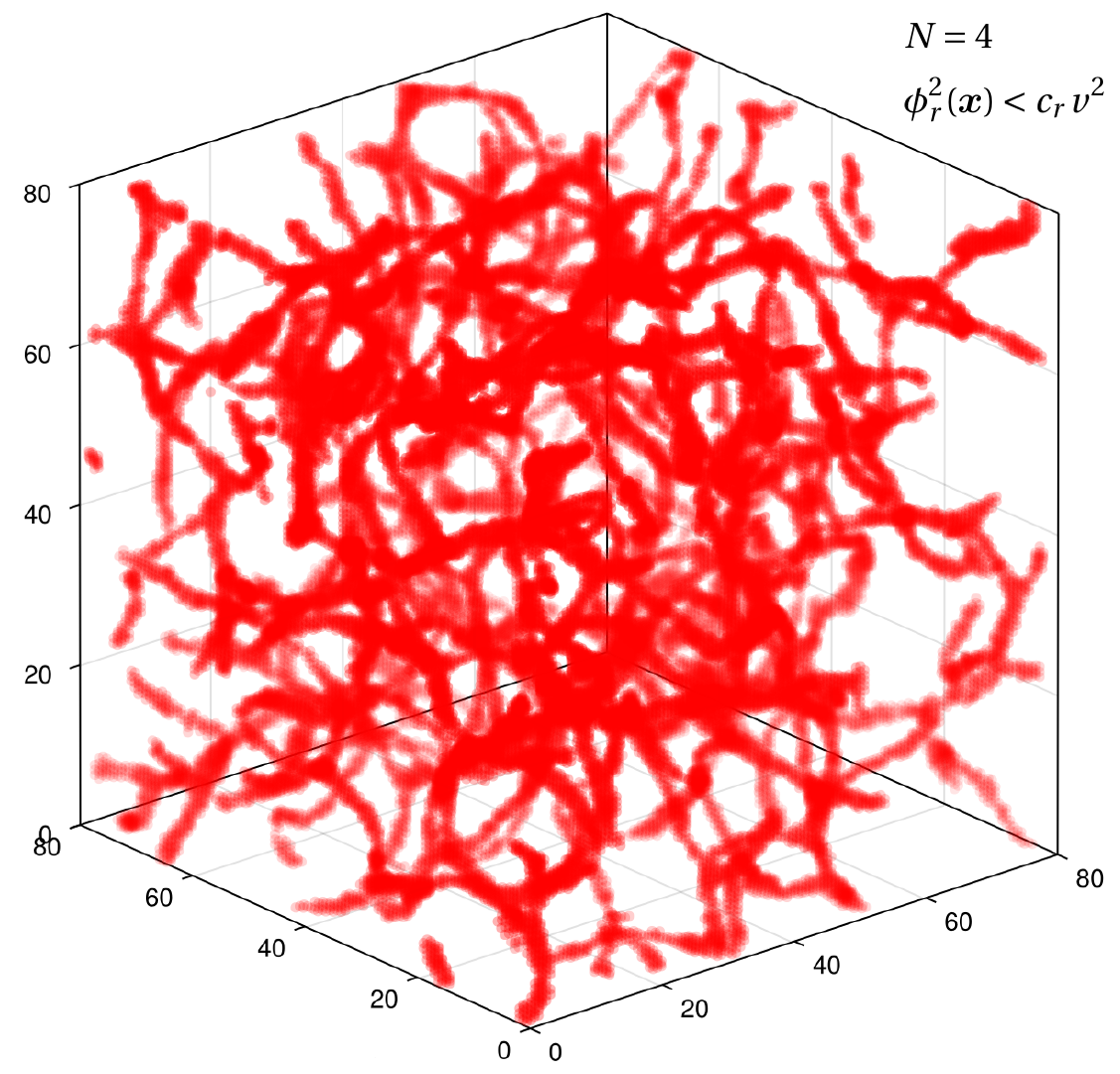}    
    \\
    \includegraphics[width=0.34\textwidth]{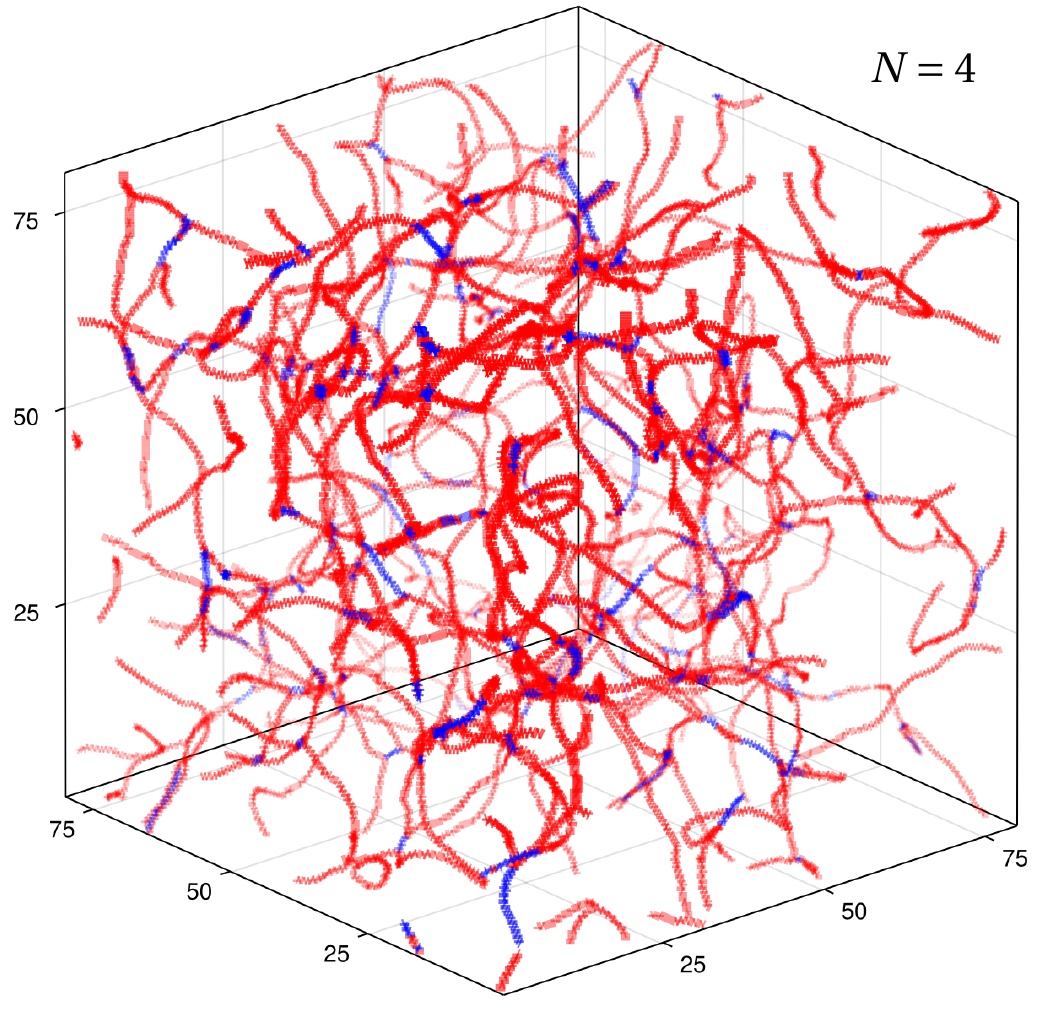}    
    \caption{
    Same as Fig.~\ref{fig:snapshot1}, but for $N=4$.  Red segments
    denote the $q=1,3$ unoriented class, while blue segments denote
    the self-conjugate $q=2$ class.  The coexistence of red and blue
    segments illustrates the charge structure that allows
    charge-conserving $1+1\leftrightarrow 2$ junctions.}
    \label{fig:snapshot2}
\end{figure}

\begin{figure}[t]
    \centering
    \includegraphics[width=0.34\textwidth]{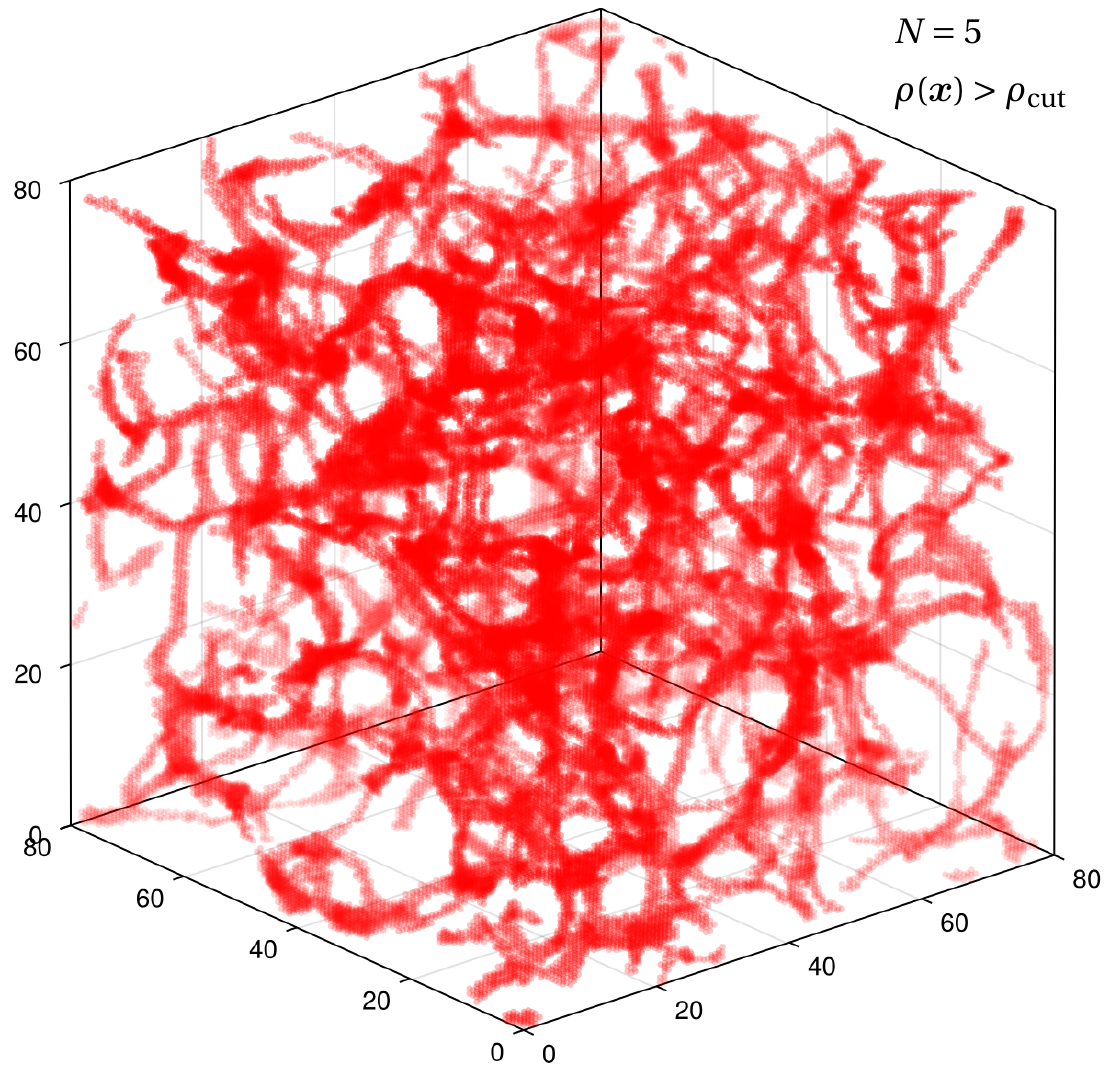}
    \\
    \includegraphics[width=0.34\textwidth]{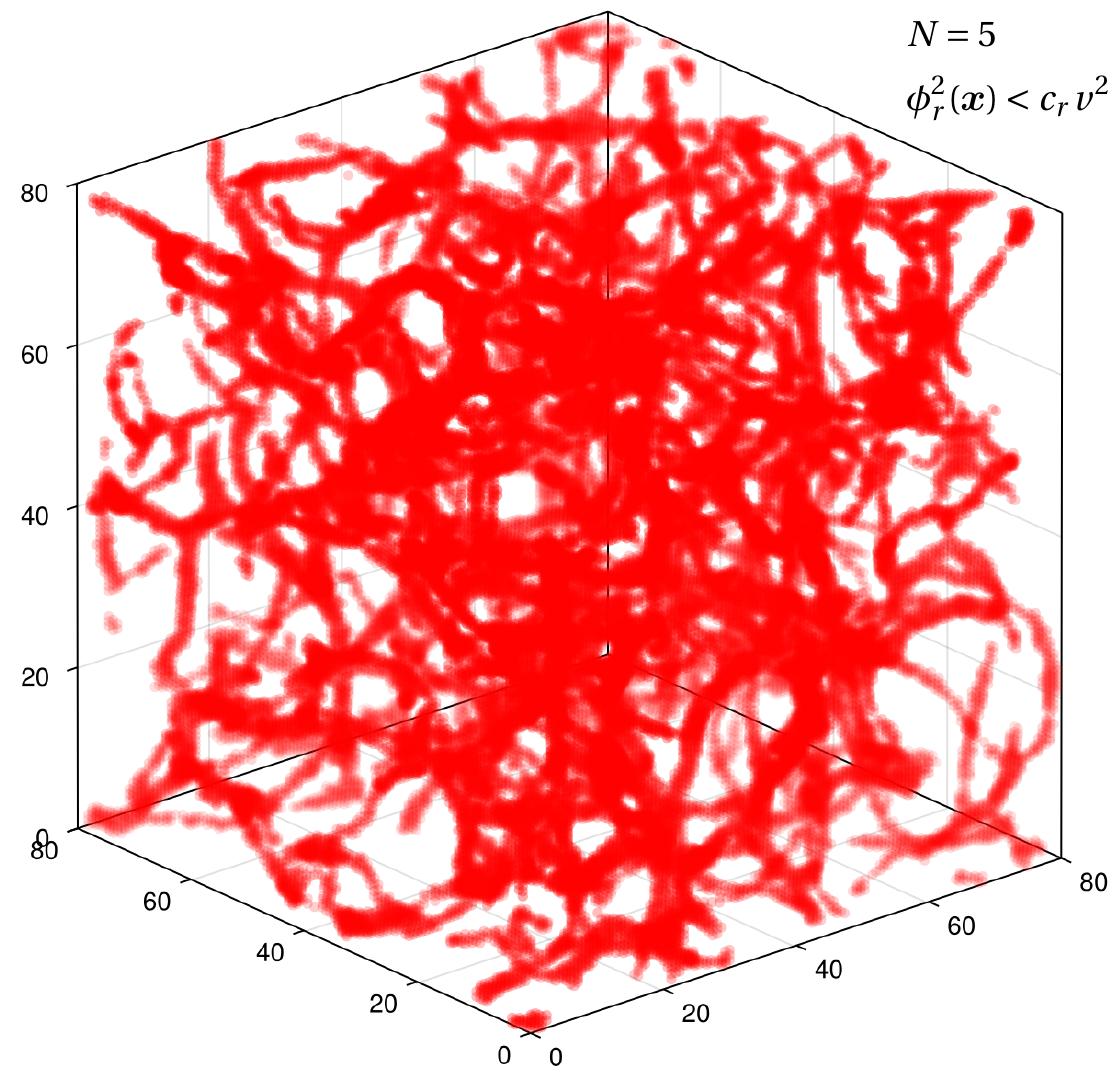}    
    \\
    \includegraphics[width=0.34\textwidth]{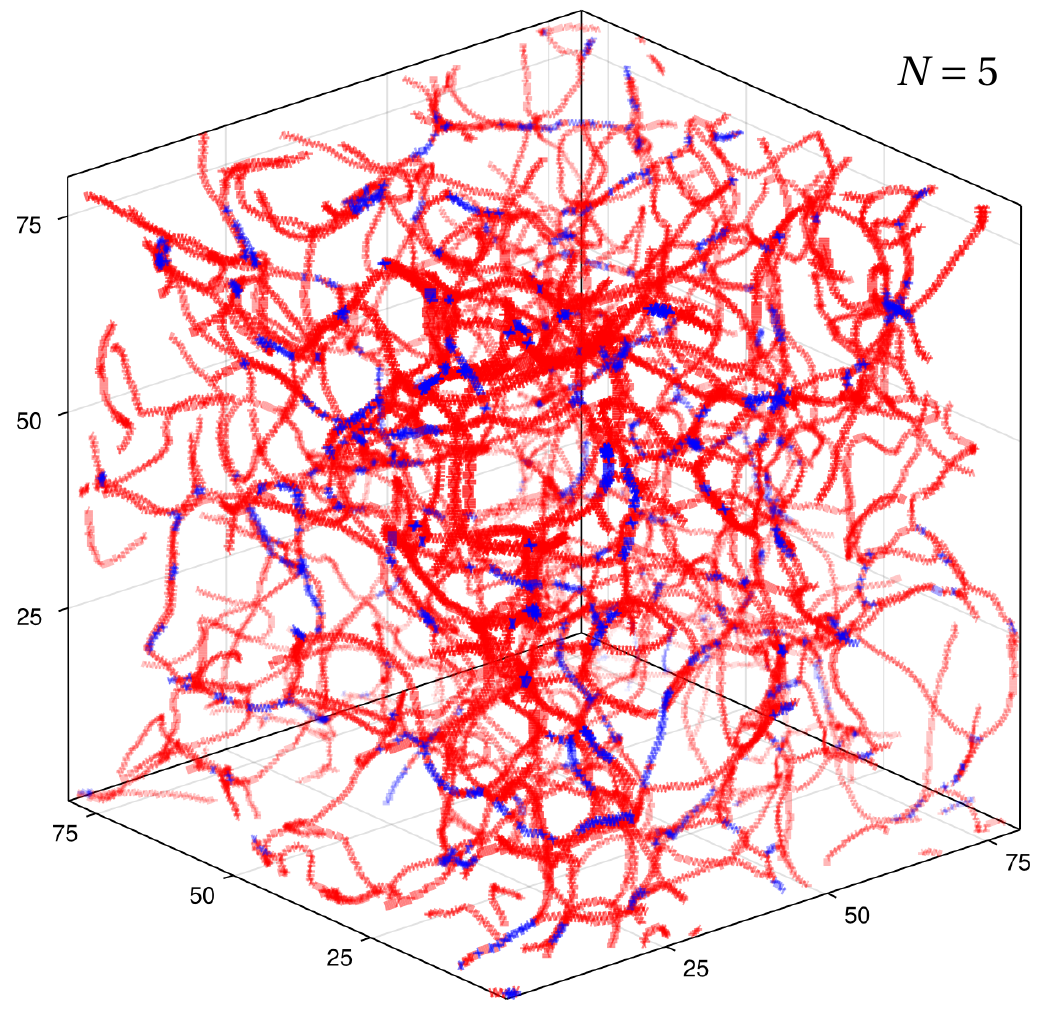}    
    \caption{
    Same as Fig.~\ref{fig:snapshot1}, but for $N=5$.  Red denotes
    the $q=1,4$ class and blue denotes the $q=2,3$ class.  The
    bottom panel shows that the non-minimal charge class is present but
    remains subdominant in the network.}
    \label{fig:snapshot3}
\end{figure}

For illustration, we show snapshots of the string network for $N=3,4$, and 5 in
Figs.~\ref{fig:snapshot1}--\ref{fig:snapshot3}.  These snapshots are
obtained from smaller runs with $L=80$, $N_L=256$, and $s=0$,
which are sufficient for visualizing the network.  In each
figure, the top panel shows high-energy regions satisfying
\begin{equation}
  \rho(\bm{x})>\rho_{\rm cut} \equiv c_E\,a^{2s-2} V(\phi=0)
\end{equation}
with $c_E=0.1$.
Neither this threshold nor an energy-density weight enters the definitions of $\zeta$ and $R_q$, which use plaquette windings.
The middle panel shows the radial-core parameter,
\begin{equation}
 \phi_r^2(\bm{x}) < c_r v^2,
\end{equation}
with $c_r=0.80,\,0.85,$ and $0.87$ for $N=3,4,$ and 5,
respectively.  
Note again that in our model the string core does not necessarily force $\phi_r^2 = 0$. 
This effect becomes more visible as $N$ increases, as shown in
Table~\ref{tab:radial-profile-core}.

The bottom panel shows the charge-resolved string
segments reconstructed from the nonzero plaquette flux $Q_p$ in
Eq.~\eqref{eq:plaquette-charge}.  Red denotes the minimal unoriented
charge class $q=1$ or $N-1$, while blue denotes the $q=2$ class
for $N=4$ and the $q=2,3$ class for $N=5$.  

The segment plots
show that non-minimal charge strings appear, and the network is not
visually dominated by highly connected many-leg junctions.  In
particular, the snapshots are consistent with charge-conserving
three-pronged rearrangements such as $1+1\leftrightarrow 2$, rather
than with a web of many-branched vertices. 
Note that the logarithmic part of
the global-string tension scales approximately as
$\mu_q\propto q(N-q)$, so that for $N=4$ one has
$\mu_2/\mu_1\simeq4/3<2$.  Two unit strings can therefore energetically
prefer to combine into a single $q=2$ string.  A four-pronged
unit-charge vertex is topologically allowed, but it need not be a generic
stable local configuration.  It can split into two three-pronged junctions
connected by a short $q=2$ segment.

\begin{figure}[t]
    \centering
    \includegraphics[width=0.45\textwidth]{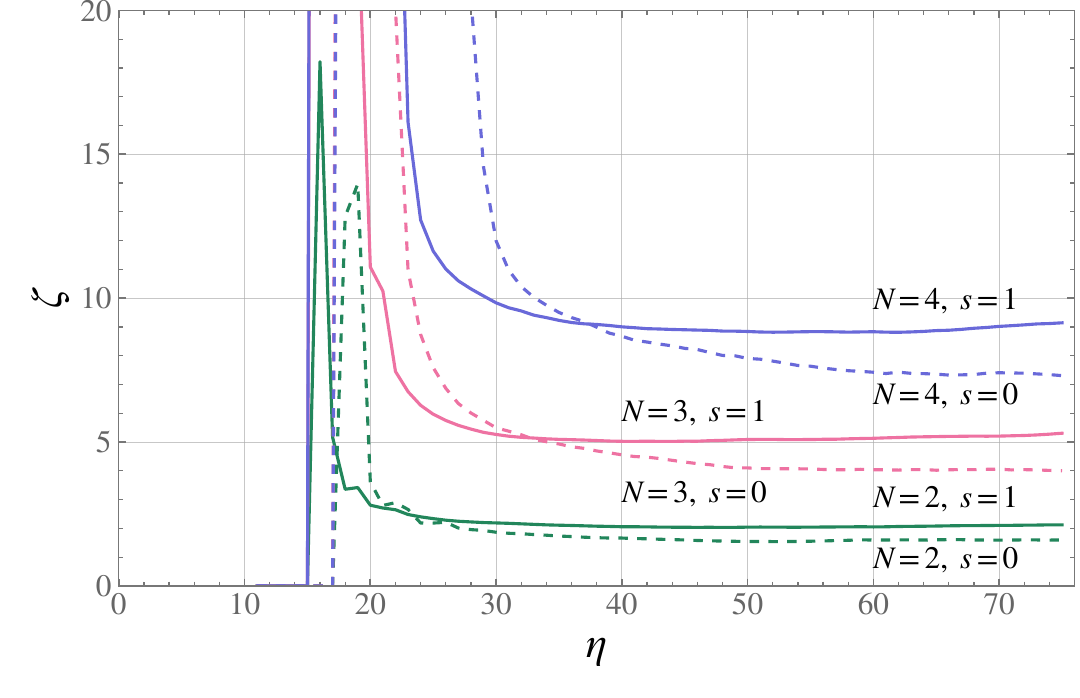}
    \\
    \includegraphics[width=0.45\textwidth]{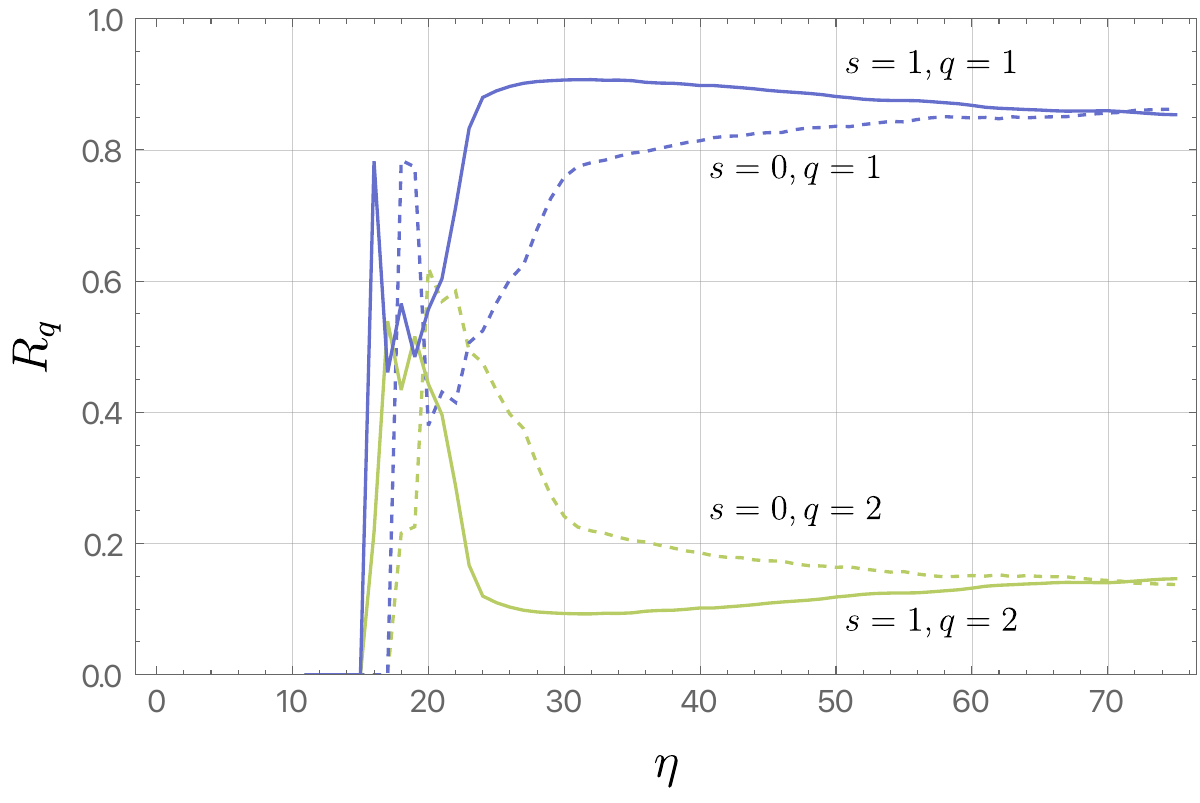}
    \caption{
    Scaling parameters as functions of conformal time.
    Top: the total scaling parameter $\zeta$ for $N=2,3,$ and $4$.
    Bottom: charge fractions $R_q$ for
    $N=4$, comparing the minimal $q=1,3$ class with the
    self-conjugate $q=2$ class. }
    \label{fig:scaling}
\end{figure}

The primary scaling observable is $\zeta(\eta)$, defined from the topological winding length in Eq.~(\ref{eq:zeta-winding}).
Figure~\ref{fig:scaling} shows this quantity and the charge fractions $R_q$ in the larger runs.
In the top panel, the green, magenta, and blue curves correspond to $N=2,3,$ and $4$, respectively.
The solid and dashed curves denote $s=1$ and $s=0$, respectively.
For $s=0$, $\zeta$ decreases toward a plateau, with a residual decrease for $N=4$ over $50\leq\eta\leq70$.
For $s=1$, $\zeta$ varies weakly after $\eta\simeq30$, with an increase of only about $2$--$3\%$ from $\eta=50$ to $70$ for $N=3,4$.
These time series alone do not establish the logarithmic growth discussed for global strings~\cite{Yamaguchi:1999yp,Fleury:2015aca,Hindmarsh:2017qff,Gorghetto:2018myk,Hindmarsh:2019csc}.

The bottom panel shows the charge fractions for $N=4$.  The minimal
unoriented charge class $q=1,3$ dominates the total length, while the
self-conjugate $q=2$ component survives at the 10--20\% level.  This
is consistent with the blue segments visible in
Fig.~\ref{fig:snapshot2}.

\begin{figure}[t]
    \centering
    \includegraphics[width=0.48\textwidth]{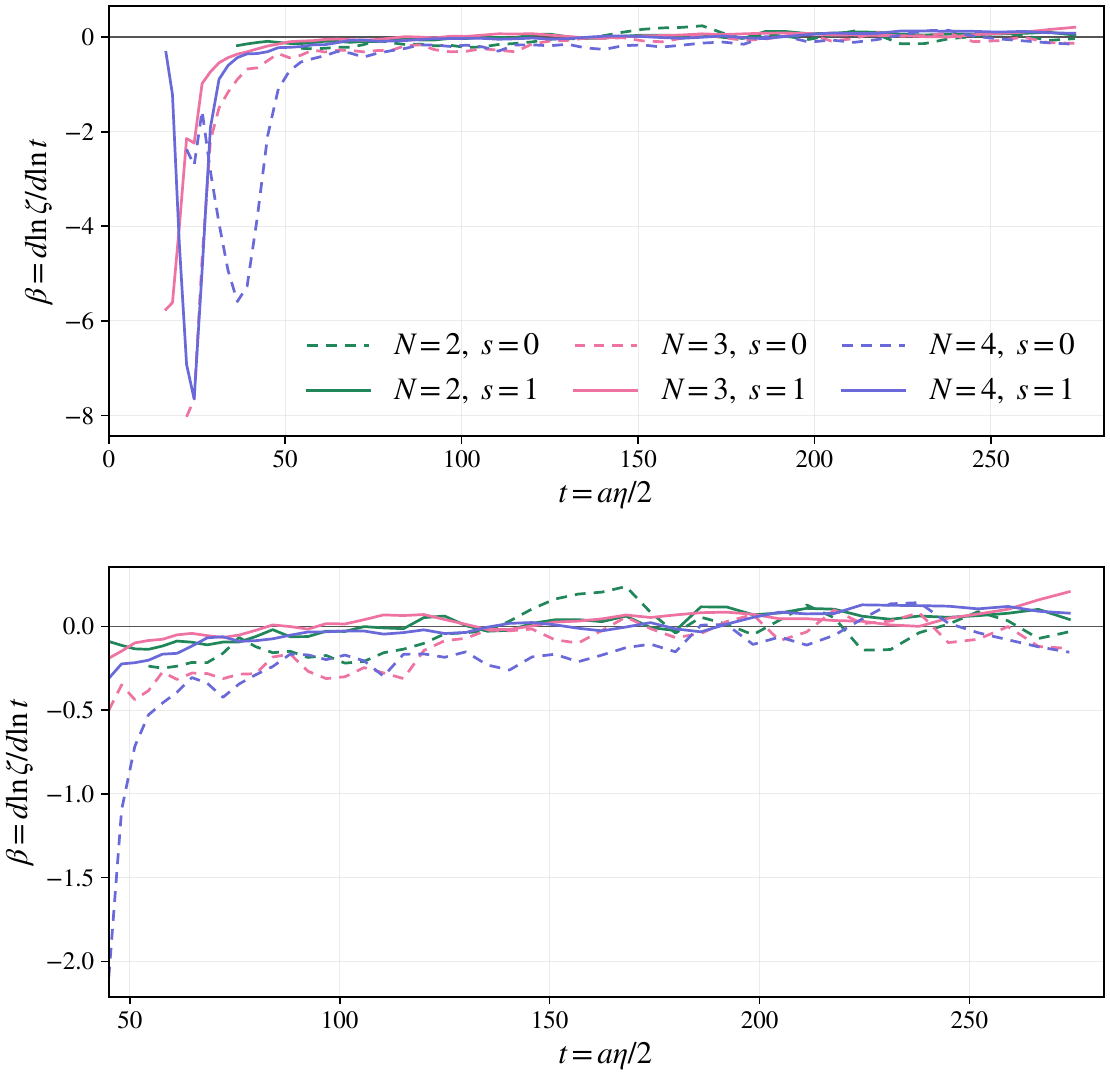}
    \caption{Logarithmic slope $\beta$, defined in Eq.~(\ref{eq:scaling-slope}), as a function of cosmic age $t=t_{\rm phys}=a\eta/2$, using the same winding-length data as the top panel of Fig.~\ref{fig:scaling}.
    Green, magenta, and blue curves correspond to $N=2,3,$ and $4$, respectively.
    Solid and dashed curves denote $s=1$ and $s=0$, respectively.
    The lower panel enlarges the late-time region.
    The horizontal line at $\beta=0$ indicates constant $\zeta$.}
    \label{fig:beta}
\end{figure}

To quantify the time dependence of $\zeta$, we define the logarithmic slope
\begin{align}
  \beta\equiv\frac{\dd\ln\zeta}{\dd\ln t_{\rm phys}}
  =\frac{1}{2}\frac{\dd\ln\zeta}{\dd\ln\eta},
  \label{eq:scaling-slope}
\end{align}
where $t_{\rm phys}=a\eta/2\propto\eta^2$ is the cosmic age in radiation domination.
Density scaling corresponds to $\beta=0$, while positive and negative values indicate an increasing and decreasing $\zeta$, respectively.
Figure~\ref{fig:beta} shows this diagnostic for the same data as the top panel of Fig.~\ref{fig:scaling}.
The initially large negative slopes become smaller as the networks evolve.
At late times, the $s=1$ curves lie close to zero with small positive excursions.
For $s=0$, the $N=3,4$ curves retain a weak, predominantly negative slope, while the $N=2$ curve fluctuates around zero with both positive and negative values.
This behavior suggests an approach toward approximate density scaling over the simulated interval.

The first seven columns of Table~\ref{tab:scaling} summarize the late-time results at $\eta=70$.
They include $N=2,3,4,5,$ and 8, with both $s=0$ and $s=1$ except for the $N=8$ row.
For $N=3$ and $s=0$, the quoted uncertainty is the standard error of the mean from 10 independent random initial conditions.
The other entries in these columns are from single production runs for each value of $s$.
Because of the computational cost, 
the time series for $N=5$ and $N=8$ are not shown in Fig.~\ref{fig:scaling}. 
Their values in Table~\ref{tab:scaling} are included for reference. 
The $N=8$ row was obtained on a $512^3$ lattice.

The last column of Table~\ref{tab:scaling} gives additional results at $\eta=50$ from simulations with $N_L=688$, $L=100$, $\Delta\eta=0.02$, $\eta_0=10$, $a(\eta_0)=a_0=1$, and an initial fluctuation amplitude of $0.1$.
We average over five random initial conditions for $N=3$, $s=1$ and three for each of the other $N=2,3,4$ ensembles.
The quoted errors are the standard errors of the means, obtained by dividing the sample standard deviation by the square root of the number of realizations.
These are values at a fixed conformal time, not fits to an asymptotic plateau.

The charge fractions are provided as reference values from a single realization for each parameter set.
For $N=4$ and $N=5$, the non-minimal charge class contributes roughly $12$--$14\%$ of the total winding length at $\eta=70$.
For $N=3$, the charge $q=2$ is the conjugate of $q=1$ and is not
listed as an independent unoriented type.  For $N=8$, the
higher-charge classes are individually at the few-percent level.

\begin{table*}[t]
\centering
\caption{Late-time scaling observables extracted from the $\ZN$ winding length.
All length densities are normalized by the square of the cosmic age $t_{\rm phys}=a\eta/2$.
The first seven columns give the results at $\eta=70$, with $\tilde{\zeta}\equiv\zeta/(N^2-1)$.
The charge fractions $R_q$ denote unoriented classes, combining charges $q$ and $N-q$ when they are conjugate.
For even $N$, the $q=N/2$ class is self-conjugate.
The last column gives the mean $\zeta$ at $\eta=50$ for $N_L=688$ and $L=100$.
Its errors are standard errors of the means from five realizations for $N=3$, $s=1$ and three for each of the other $N=2,3,4$ ensembles, without systematic uncertainties.
The errors for $N=3$, $s=0$ at $\eta=70$ are standard errors of the means from 10 realizations.
}
\setlength{\tabcolsep}{7pt}
\begin{tabular}{cccccccc}
\toprule
$N$ & $s$ & $\zeta$ & $\tilde{\zeta}$ & $R_{q=2}$ & $R_{q=3}$ & $R_{q=4}$ & $\zeta(\eta=50),\ N_L=688$\\
\midrule
2 & 0 & 1.59 & 0.530 & -- & -- & -- & $1.611 \pm 0.036$ \\
3 & 0 & $3.952 \pm 0.066$ & $0.494 \pm 0.008$ & -- & -- & -- & $4.390 \pm 0.122$ \\
4 & 0 & 7.41 & 0.494 & 14\% & -- & -- & $7.976 \pm 0.060$ \\
5 & 0 & 13.0 & 0.540 & 14\% & -- & -- & -- \\
8 & 0 & 32.1 & 0.509 & 5\% & 2\% & 2\% & -- \\
2 & 1 & 2.10 & 0.701 & -- & -- & -- & $2.050 \pm 0.039$ \\
3 & 1 & 5.21 & 0.651 & -- & -- & -- & $4.956 \pm 0.029$ \\
4 & 1 & 9.02 & 0.601 & 14\% & -- & -- & $8.722 \pm 0.144$ \\
5 & 1 & 14.2 & 0.590 & 12\% & -- & -- & -- \\
\bottomrule
\end{tabular}
\label{tab:scaling}
\end{table*}

To compare the measured $N$ dependence with the adjoint dimension $N^2-1$, we define
\begin{align}
\tilde{\zeta}\equiv \frac{\zeta}{N^2-1} .
\end{align}
We fit $\zeta=A_s(N^2-1)$ separately for $s=0$ and $s=1$, using only the $N=2,3,4$ ensembles at $\eta=50$ on the $688^3$ lattice.
Weighting by the inverse squared standard errors and treating the ensemble means as independent gives
\begin{align}
A_0&=0.5332\pm0.0037, \nonumber\\
A_1&=0.6191\pm0.0033.
\label{eq:N-dependence-fit}
\end{align}
The coefficient errors are formal statistical uncertainties based on the quoted standard errors.
The values in Table~\ref{tab:scaling} and the coefficients $A_s$ are obtained at the quoted times for the initial conditions adopted in this study.
For $s=0$, the data are consistent with this proportionality within the quoted errors.
For $s=1$, the deviations from the fit are $+10.4\%$, $+0.1\%$, and $-6.1\%$ for $N=2,3,4$, respectively.
Thus the proportionality describes the $s=1$ trend at approximately the $10\%$ level, but does not fit the data within their statistical errors.

The single-run results at $\eta=70$ in Table~\ref{tab:scaling} differ from these fits by at most about $8\%$ for $s=0$ and $13\%$ for $s=1$.
For $s=1$ and $N\ge3$, the differences are about $5\%$ or less.
The $N=3$, $s=0$ ensemble on the $1024^3$ lattice at $\eta=70$, which is not included in the fit, lies $7.4\%$ below the fitted value and $10.0\%$ below the $688^3$ ensemble mean at $\eta=50$.
The different comparison times and lattice parameters prevent attributing these differences solely to spatial resolution.
The proportionality is an approximate empirical relation for the parameters studied.
The comparison holds $m$, $g$, and the curvature at the origin fixed, while $v$ and $\lambda_r$ vary with $N$ as specified above.

\subsection{Implications for gravitational-wave estimates}
\label{subsec:gw-implications}

The simulations estimate the network-density input to gravitational-wave calculations in this effective global model.
The physical length density of strings with charge $q$ is
\begin{align}
  \frac{\ell_{{\rm com},q}}{a^2 L^3}
  =\frac{\zeta_q}{t_{\rm phys}^2}.
\end{align}
Using charge-dependent tensions $\mu_q$ that include the logarithmic gradient energy at a chosen infrared cutoff, and neglecting velocity corrections and junction energies, gives the parametric estimate
\begin{align}
  \frac{\rho_{\rm str}}{\rho_{\rm tot}}
  \sim G\sum_q\mu_q\zeta_q
  =G\mu_{\rm eff}\zeta,
  \label{eq:string-energy-fraction}
\end{align}
up to order-one factors in the radiation era, where
\begin{align}
  \mu_{\rm eff}\equiv\sum_q R_q\mu_q.
  \label{eq:effective-tension}
\end{align}
Thus $\zeta=\tilde\zeta(N^2-1)$ enhances this density input by $N^2-1$ at fixed $\mu_{\rm eff}$ and $\tilde\zeta$.
In calculations of gravitational waves from loops, the abundance enters through the loop number distribution~\cite{Ringeval:2017eww}, while Goldstone radiation affects the lifetimes of global-string loops~\cite{Chang:2021afa}.

%%%%%%%%%%%%%%%%%%%%%%%%%%%%%%%%%%%%%%%%%%%%%%%%%%%
\section{Discussion and conclusions}
\label{sec:discussion}
%%%%%%%%%%%%%%%%%%%%%%%%%%%%%%%%%%%%%%%%%%%%%%%%%%%

We have formulated a scalar-only model with vacuum manifold
$\PSU(N)=\SU(N)/\ZN$ and developed a numerical method for studying
the formation and evolution of global $\ZN$ string networks in a
radiation-dominated universe.  The model contains three adjoint scalar
fields and is motivated by the irreducible $\SU(2)$ embedding vacuum
of the mass-deformed $\mathcal{N}=4$ Yang--Mills theory.  Because the
center $\ZN$ acts trivially on adjoint fields, the faithful global
symmetry is $\PSU(N)$, and the vacuum manifold has
$\pi_1(\PSU(N))=\ZN$.

Strings of this type can meet at a vertex joining $N$ unit-charge strings.
This should be distinguished from non-Abelian string networks in which the
junctions cannot disappear unless the network is untied.  Such networks are
expected to become frustrated and can dominate the energy density of the
Universe~\cite{Spergel:1996ai,McGraw:1997nx,Avgoustidis:2007aa}.  By contrast, in the present $\ZN$ string system, vertices can
disappear, for example by annihilating with other vertices.

We calculate the total winding-length density $\zeta$ from plaquette windings.
For $N=2,3,$ and $4$, the time histories suggest an approach toward approximate density scaling over the simulated interval.
For $N=5$ and $8$, we report densities and charge fractions at the specified times.
The non-minimal charge components remain subdominant, while the normalization of $\zeta$ grows approximately in proportion to $N^2-1$ for the parameters studied.

Our numerical simulations are limited to $N\le 8$.  For larger $N$, strings
with larger charge may become more important.  However, the number of
scalar fields scales as $N^2-1$, so the numerical cost grows rapidly with
$N$.  As the winding-length density increases, resolving separated string cores within a Hubble volume becomes more demanding.
This requirement increases the numerical cost further.  Investigating larger $N$ is left for future work.

The scalar-only model describes global $\ZN$ strings, rather than local flux
tubes of the dual $\PSU(N)$ gauge theory.  Gauge fields would screen
the angular gradient energy, whereas in the scalar-only model the
string tension receives an infrared logarithm.  
A direct treatment of local flux tubes, baryon vertices, and their reconnection probabilities remains an important direction for future work.
The present $N=2,3,4$ time histories suggest that the junction structure associated with $\ZN$ charge conservation does not by itself prevent an approach toward approximate density scaling over the simulated interval.

\section*{Acknowledgements}
This work was supported by JSPS KAKENHI Grant Number 23K13092.
Some of the numerical computations in this work were performed using the supercomputing resources provided by the Cyberscience Center, Tohoku University.

\appendix

\section{Mass spectrum around the Higgs vacuum}
\label{sec:appA}

In this Appendix, we explain the calculation of the curvature of the potential, $\partial^2 V/\partial\phi^A\partial\phi^B$, at the Higgs vacuum.

To see the eigenvalues explicitly, we write
\begin{equation}
  \Phi_i=\Phi_i^{({\rm vac})}+\delta\Phi_i .
\end{equation}
The linearized $F$-term is
\begin{equation}
  \delta F_i^a T^a
  =
  m\,\delta\Phi_i
  +i m\,\epsilon_{ijk}\,[\delta\Phi_j,t_k].
  \label{eq:linearized-Fterm}
\end{equation}

Let $V_j$ be the $N$-dimensional irreducible $\SU(2)$ representation, with $j=(N-1)/2$.  The adjoint field transforms as
\begin{equation}
  V_j\otimes V_j^\ast
  =\bigoplus_{\ell=0}^{2j}V_\ell, 
\end{equation}
except that 
the $\ell=0$ component is the identity matrix and is absent from $\mathfrak{su}(N)$.  Hence the adjoint representation of $\SU(N)$ decomposes under the principal $\SU(2)$ as
\begin{equation}
  \mathfrak{su}(N)
  =
  \bigoplus_{\ell=1}^{N-1}V_\ell .
  \label{eq:principal-adjoint-decomp}
\end{equation}

If $L_k$ denotes the principal adjoint action $L_k X=[t_k,X]$, and $S_k$ denotes the spin-one generator acting on the flavor index ($(S_k)_{ij}=-i\epsilon_{ijk}$), Eq.~(\ref{eq:linearized-Fterm}) can be written as
\begin{equation}
  \delta F=m\left(1+\bm S\cdot\bm L\right)\delta\Phi .
  \label{eq:Fterm-spin-operator}
\end{equation}
On a component with total spin $J$, where $\bm J=\bm S+\bm L$, one has
\begin{equation}
  \bm S\cdot\bm L
  =
  \frac12
  \left[
    J(J+1)-1(1+1)-\ell(\ell+1)
  \right].
\end{equation}
For the three Clebsch--Gordan branches $J=\ell-1,\ell,\ell+1$, the eigenvalues of $1+\bm S\cdot\bm L$ are
\begin{equation}
  -\ell,\qquad 0,\qquad \ell+1 .
  \label{eq:Fterm-linear-eigenvalues}
\end{equation}
Since the vacuum satisfies $F_i=0$, the $F$-term contribution to the Hessian is the square of this linear operator.

The middle branch in Eq.~\eqref{eq:Fterm-linear-eigenvalues} is tangent to the vacuum orbit and gives the $N^2-1$ Goldstone directions,
\begin{equation}
  \sum_{\ell=1}^{N-1}(2\ell+1)=N^2-1.
\end{equation}
The nonzero F-term Hessian eigenvalues in the nonradial directions are therefore
\begin{align}
  M_{\ell-1}^2&=  m^2\ell^2,
  \qquad \ell=2,\ldots,N-1,
\\
  M_{\ell+1}^2&= m^2(\ell+1)^2,
  \qquad \ell=1,\ldots,N-1.
\end{align}
The $\ell=1$, $J=0$ singlet is the radial direction $\delta\Phi_i\propto \Phi_i^{({\rm vac})}$.  The radial stabilizing term in Eq.~\eqref{eq:potential} contributes only along this direction at quadratic order:
\begin{equation}
  \frac{\lambda_r}{4}
  \left(\phi_r^2-v^2\right)^2
  =
  \lambda_r
  \left(\phi^{({\rm vac})}\cdot\delta\phi\right)^2
  +O(\delta\phi^3).
\end{equation}
Thus the radial singlet receives an additional $2\lambda_r v^2$ contribution, and its mass is
\begin{equation}
  M_{\rm rad}^2= m^2+2\lambda_r v^2 .
\end{equation}

\section{Topological string-length reconstruction}
\label{sec:appC}

\begin{figure}[tt]
  \centering
  \includegraphics[width=0.92\linewidth]{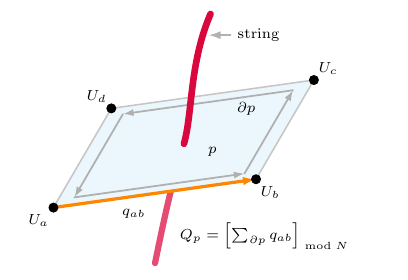}
  \caption{Schematic of the plaquette-winding diagnostic used in the
  numerical reconstruction of the string length.  At each lattice site we
  choose a local representative $U_a$ of the $\PSU(N)$ orientation.
  For an oriented link $a\to b$, the nearest-center projection defines
  $q_{ab}$.  The plaquette charge $Q_p$ is then obtained by summing
  $q_{ab}$ around the oriented boundary $\partial p$ modulo $N$.
  A nonzero $Q_p$ means that a $\ZN$ string pierces the plaquette, as
  shown by the red string passing through $p$.}
  \label{fig:plaquette-charge}
\end{figure}

In this appendix, we explain the numerical method used to identify the
plaquettes pierced by cosmic strings.  
This procedure is a center-valued version of the Vachaspati--Vilenkin
plaquette-winding construction~\cite{Vachaspati:1984dz,Strobl:1996yj}.
This reconstruction gives an
estimate of the total string length and also determines the string
charge.

The scaling analysis uses the center winding of the full field configuration.  At each lattice site away from the string core we reconstruct a local $\PSU(N)$ orientation from the three adjoint scalar fields.  Writing
\begin{equation}
  \Phi_i(\bm{x})=\phi_i^a(\bm{x})T^a,
  \qquad i=1,2,3,
\end{equation}
we determine an $SU(N)$ representative $U(\bm{x})$ such that, near the Higgs vacuum,
\begin{equation}
  \Phi_i(\bm{x})\simeq U(\bm{x}) \Phi_i^{({\rm vac})} U(\bm{x})^\dagger .
\end{equation}
The representative $U(\bm{x})$ is defined only up to multiplication by an element of the center $\ZN$, and this remaining ambiguity is precisely what is used to measure the string charge.

For each oriented link we first compute the real number
\begin{equation}
  x_{ab}
  =
  \frac{N}{2\pi}
  \arg \frac{1}{N}\Tr(U_a^\dagger U_b).
\end{equation}
If the two neighboring orientations are close to the same point in
$\PSU(N)$, $x_{ab}$ lies close to an integer.  We then define the
$\ZN$-valued link variable by projecting onto the nearest integer,
\begin{equation}
  q_{ab}
  =
  \left[\operatorname{nint}(x_{ab})\right]_{\bmod N}.
\label{eq:link-charge}
\end{equation}
The individual $q_{ab}$ depends on the choice of the $\SU(N)$ lift at
each lattice site, whereas the plaquette charge
\begin{equation}
  Q_p=\left[\sum_{\partial p}q_{ab}\right]_{\bmod N}
  \label{eq:plaquette-charge}
\end{equation}
is invariant under local center redefinitions and is used to identify
strings.  This construction is illustrated in Fig.~\ref{fig:plaquette-charge}. 
The link variables $q_{ab}$ are summed around the oriented boundary
$\partial p$, and a nonzero center winding indicates that a string pierces
the plaquette.
A plaquette with $Q_p\neq0$ is counted as being pierced by a $\ZN$ string of charge $Q_p$.

Reversing the orientation of the pierced plaquette maps $q$ to $N-q$.  We
therefore group conjugate charges into unoriented string types and denote
by $n_q$ the number of pierced plaquettes with charge $q=1,\ldots,N/2$ for
even $N$ and $q=1,\ldots,(N-1)/2$ for odd $N$.

For an isotropic string network, the plaquette-piercing count estimates the Manhattan length, namely a length measured along lattice axes.  We convert it to an approximate Euclidean comoving length by
\begin{equation}
  \ell_{{\rm com},q}=\frac{2}{3}\Delta x\, n_q.
  \label{eq:app-winding-length}
\end{equation}
The factor $2/3$ is the standard isotropic correction from Manhattan length to Euclidean length.

\bibliography{reference}

\providecommand{\href}[2]{#2}\begingroup\raggedright\begin{thebibliography}{10}

\bibitem{Kibble:1976sj}
T.~W.~B. Kibble, \emph{{Topology of Cosmic Domains and Strings}}, \href{https://doi.org/10.1088/0305-4470/9/8/029}{\emph{J. Phys. A} {\bfseries 9} (1976) 1387}.

\bibitem{Vilenkin:2000jqa}
A.~Vilenkin and E.~P.~S. Shellard, \emph{{Cosmic Strings and Other Topological Defects}}. Cambridge University Press, 7, 2000.

\bibitem{Bennett:1987vf}
D.~P. Bennett and F.~R. Bouchet, \emph{{Evidence for a Scaling Solution in Cosmic String Evolution}}, \href{https://doi.org/10.1103/PhysRevLett.60.257}{\emph{Phys. Rev. Lett.} {\bfseries 60} (1988) 257}.

\bibitem{Allen:1990tv}
B.~Allen and E.~P.~S. Shellard, \emph{{Cosmic string evolution: a numerical simulation}}, \href{https://doi.org/10.1103/PhysRevLett.64.119}{\emph{Phys. Rev. Lett.} {\bfseries 64} (1990) 119}.

\bibitem{Vilenkin:1981bx}
A.~Vilenkin, \emph{{Gravitational radiation from cosmic strings}}, \href{https://doi.org/10.1016/0370-2693(81)91144-8}{\emph{Phys. Lett. B} {\bfseries 107} (1981) 47}.

\bibitem{Vachaspati:1984gt}
T.~Vachaspati and A.~Vilenkin, \emph{{Gravitational Radiation from Cosmic Strings}}, \href{https://doi.org/10.1103/PhysRevD.31.3052}{\emph{Phys. Rev. D} {\bfseries 31} (1985) 3052}.

\bibitem{NANOGrav:2023gor}
{\scshape NANOGrav} collaboration, G.~Agazie et~al., \emph{{The NANOGrav 15 yr Data Set: Evidence for a Gravitational-wave Background}}, \href{https://doi.org/10.3847/2041-8213/acdac6}{\emph{Astrophys. J. Lett.} {\bfseries 951} (2023) L8} [\href{https://arxiv.org/abs/2306.16213}{{\ttfamily 2306.16213}}].

\bibitem{EPTA:2023fyk}
{\scshape EPTA, InPTA:} collaboration, J.~Antoniadis et~al., \emph{{The second data release from the European Pulsar Timing Array - III. Search for gravitational wave signals}}, \href{https://doi.org/10.1051/0004-6361/202346844}{\emph{Astron. Astrophys.} {\bfseries 678} (2023) A50} [\href{https://arxiv.org/abs/2306.16214}{{\ttfamily 2306.16214}}].

\bibitem{Reardon:2023gzh}
D.~J. Reardon et~al., \emph{{Search for an Isotropic Gravitational-wave Background with the Parkes Pulsar Timing Array}}, \href{https://doi.org/10.3847/2041-8213/acdd02}{\emph{Astrophys. J. Lett.} {\bfseries 951} (2023) L6} [\href{https://arxiv.org/abs/2306.16215}{{\ttfamily 2306.16215}}].

\bibitem{Xu:2023wog}
H.~Xu et~al., \emph{{Searching for the Nano-Hertz Stochastic Gravitational Wave Background with the Chinese Pulsar Timing Array Data Release I}}, \href{https://doi.org/10.1088/1674-4527/acdfa5}{\emph{Res. Astron. Astrophys.} {\bfseries 23} (2023) 075024} [\href{https://arxiv.org/abs/2306.16216}{{\ttfamily 2306.16216}}].

\bibitem{NANOGrav:2023hvm}
{\scshape NANOGrav} collaboration, A.~Afzal et~al., \emph{{The NANOGrav 15 yr Data Set: Search for Signals from New Physics}}, \href{https://doi.org/10.3847/2041-8213/acdc91}{\emph{Astrophys. J. Lett.} {\bfseries 951} (2023) L11} [\href{https://arxiv.org/abs/2306.16219}{{\ttfamily 2306.16219}}].

\bibitem{Yamada:2022imq}
M.~Yamada and K.~Yonekura, \emph{{Cosmic strings from pure Yang{\textendash}Mills theory}}, \href{https://doi.org/10.1103/PhysRevD.106.123515}{\emph{Phys. Rev. D} {\bfseries 106} (2022) 123515} [\href{https://arxiv.org/abs/2204.13123}{{\ttfamily 2204.13123}}].

\bibitem{Yamada:2022aax}
M.~Yamada and K.~Yonekura, \emph{{Cosmic F- and D-strings from pure Yang{\textendash}Mills theory}}, \href{https://doi.org/10.1016/j.physletb.2023.137724}{\emph{Phys. Lett. B} {\bfseries 838} (2023) 137724} [\href{https://arxiv.org/abs/2204.13125}{{\ttfamily 2204.13125}}].

\bibitem{Yamada:2023thl}
M.~Yamada and K.~Yonekura, \emph{{Dark baryon from pure Yang-Mills theory and its GW signature from cosmic strings}}, \href{https://doi.org/10.1007/JHEP09(2023)197}{\emph{JHEP} {\bfseries 09} (2023) 197} [\href{https://arxiv.org/abs/2307.06586}{{\ttfamily 2307.06586}}].

\bibitem{Witten:1985fp}
E.~Witten, \emph{{Cosmic Superstrings}}, \href{https://doi.org/10.1016/0370-2693(85)90540-4}{\emph{Phys. Lett. B} {\bfseries 153} (1985) 243}.

\bibitem{Polchinski:1988cn}
J.~Polchinski, \emph{{Collision of Macroscopic Fundamental Strings}}, \href{https://doi.org/10.1016/0370-2693(88)90942-2}{\emph{Phys. Lett. B} {\bfseries 209} (1988) 252}.

\bibitem{Dvali:2003zj}
G.~Dvali and A.~Vilenkin, \emph{{Formation and evolution of cosmic D strings}}, \href{https://doi.org/10.1088/1475-7516/2004/03/010}{\emph{JCAP} {\bfseries 03} (2004) 010} [\href{https://arxiv.org/abs/hep-th/0312007}{{\ttfamily hep-th/0312007}}].

\bibitem{Copeland:2003bj}
E.~J. Copeland, R.~C. Myers and J.~Polchinski, \emph{{Cosmic F and D strings}}, \href{https://doi.org/10.1088/1126-6708/2004/06/013}{\emph{JHEP} {\bfseries 06} (2004) 013} [\href{https://arxiv.org/abs/hep-th/0312067}{{\ttfamily hep-th/0312067}}].

\bibitem{Jackson:2004zg}
M.~G. Jackson, N.~T. Jones and J.~Polchinski, \emph{{Collisions of cosmic F and D-strings}}, \href{https://doi.org/10.1088/1126-6708/2005/10/013}{\emph{JHEP} {\bfseries 10} (2005) 013} [\href{https://arxiv.org/abs/hep-th/0405229}{{\ttfamily hep-th/0405229}}].

\bibitem{Hanany:2005bc}
A.~Hanany and K.~Hashimoto, \emph{{Reconnection of colliding cosmic strings}}, \href{https://doi.org/10.1088/1126-6708/2005/06/021}{\emph{JHEP} {\bfseries 06} (2005) 021} [\href{https://arxiv.org/abs/hep-th/0501031}{{\ttfamily hep-th/0501031}}].

\bibitem{Ellis:2023tsl}
J.~Ellis, M.~Lewicki, C.~Lin and V.~Vaskonen, \emph{{Cosmic superstrings revisited in light of NANOGrav 15-year data}}, \href{https://doi.org/10.1103/PhysRevD.108.103511}{\emph{Phys. Rev. D} {\bfseries 108} (2023) 103511} [\href{https://arxiv.org/abs/2306.17147}{{\ttfamily 2306.17147}}].

\bibitem{Douglas:1995nw}
M.~R. Douglas and S.~H. Shenker, \emph{{Dynamics of SU(N) supersymmetric gauge theory}}, \href{https://doi.org/10.1016/0550-3213(95)00258-T}{\emph{Nucl. Phys. B} {\bfseries 447} (1995) 271} [\href{https://arxiv.org/abs/hep-th/9503163}{{\ttfamily hep-th/9503163}}].

\bibitem{Hanany:1997hr}
A.~Hanany, M.~J. Strassler and A.~Zaffaroni, \emph{{Confinement and strings in MQCD}}, \href{https://doi.org/10.1016/S0550-3213(97)00651-2}{\emph{Nucl. Phys. B} {\bfseries 513} (1998) 87} [\href{https://arxiv.org/abs/hep-th/9707244}{{\ttfamily hep-th/9707244}}].

\bibitem{Witten:1998zw}
E.~Witten, \emph{{Anti-de Sitter space, thermal phase transition, and confinement in gauge theories}}, \href{https://doi.org/10.4310/ATMP.1998.v2.n3.a3}{\emph{Adv. Theor. Math. Phys.} {\bfseries 2} (1998) 505} [\href{https://arxiv.org/abs/hep-th/9803131}{{\ttfamily hep-th/9803131}}].

\bibitem{Polchinski:2000uf}
J.~Polchinski and M.~J. Strassler, \emph{{The String dual of a confining four-dimensional gauge theory}},  \href{https://arxiv.org/abs/hep-th/0003136}{{\ttfamily hep-th/0003136}}.

\bibitem{Klebanov:2000hb}
I.~R. Klebanov and M.~J. Strassler, \emph{{Supergravity and a confining gauge theory: Duality cascades and chi SB resolution of naked singularities}}, \href{https://doi.org/10.1088/1126-6708/2000/08/052}{\emph{JHEP} {\bfseries 08} (2000) 052} [\href{https://arxiv.org/abs/hep-th/0007191}{{\ttfamily hep-th/0007191}}].

\bibitem{Maldacena:2000yy}
J.~M. Maldacena and C.~Nunez, \emph{{Towards the large N limit of pure N=1 superYang-Mills}}, \href{https://doi.org/10.1103/PhysRevLett.86.588}{\emph{Phys. Rev. Lett.} {\bfseries 86} (2001) 588} [\href{https://arxiv.org/abs/hep-th/0008001}{{\ttfamily hep-th/0008001}}].

\bibitem{Vafa:2000wi}
C.~Vafa, \emph{{Superstrings and topological strings at large N}}, \href{https://doi.org/10.1063/1.1376161}{\emph{J. Math. Phys.} {\bfseries 42} (2001) 2798} [\href{https://arxiv.org/abs/hep-th/0008142}{{\ttfamily hep-th/0008142}}].

\bibitem{Witten:1998xy}
E.~Witten, \emph{{Baryons and branes in anti-de Sitter space}}, \href{https://doi.org/10.1088/1126-6708/1998/07/006}{\emph{JHEP} {\bfseries 07} (1998) 006} [\href{https://arxiv.org/abs/hep-th/9805112}{{\ttfamily hep-th/9805112}}].

\bibitem{Spergel:1996ai}
D.~Spergel and U.-L. Pen, \emph{{Cosmology in a string dominated universe}}, \href{https://doi.org/10.1086/311074}{\emph{Astrophys. J. Lett.} {\bfseries 491} (1997) L67} [\href{https://arxiv.org/abs/astro-ph/9611198}{{\ttfamily astro-ph/9611198}}].

\bibitem{McGraw:1997nx}
P.~McGraw, \emph{{Evolution of a nonAbelian cosmic string network}}, \href{https://doi.org/10.1103/PhysRevD.57.3317}{\emph{Phys. Rev. D} {\bfseries 57} (1998) 3317} [\href{https://arxiv.org/abs/astro-ph/9706182}{{\ttfamily astro-ph/9706182}}].

\bibitem{Avgoustidis:2007aa}
A.~Avgoustidis and E.~P.~S. Shellard, \emph{{Velocity-Dependent Models for Non-Abelian/Entangled String Networks}}, \href{https://doi.org/10.1103/PhysRevD.78.103510}{\emph{Phys. Rev. D} {\bfseries 78} (2008) 103510} [\href{https://arxiv.org/abs/0705.3395}{{\ttfamily 0705.3395}}].

\bibitem{Vachaspati:1986cc}
T.~Vachaspati and A.~Vilenkin, \emph{{Evolution of cosmic networks}}, \href{https://doi.org/10.1103/PhysRevD.35.1131}{\emph{Phys. Rev. D} {\bfseries 35} (1987) 1131}.

\bibitem{Ng:2008mp}
Y.~Ng, T.~W.~B. Kibble and T.~Vachaspati, \emph{{Formation of Non-Abelian Monopoles Connected by Strings}}, \href{https://doi.org/10.1103/PhysRevD.78.046001}{\emph{Phys. Rev. D} {\bfseries 78} (2008) 046001} [\href{https://arxiv.org/abs/0806.0155}{{\ttfamily 0806.0155}}].

\bibitem{Copeland:2005cy}
E.~J. Copeland and P.~M. Saffin, \emph{{On the evolution of cosmic-superstring networks}}, \href{https://doi.org/10.1088/1126-6708/2005/11/023}{\emph{JHEP} {\bfseries 11} (2005) 023} [\href{https://arxiv.org/abs/hep-th/0505110}{{\ttfamily hep-th/0505110}}].

\bibitem{Hindmarsh:2006qn}
M.~Hindmarsh and P.~M. Saffin, \emph{{Scaling in a SU(2)$/\mathbb{Z}_{3}$ model of cosmic superstring networks}}, \href{https://doi.org/10.1088/1126-6708/2006/08/066}{\emph{JHEP} {\bfseries 08} (2006) 066} [\href{https://arxiv.org/abs/hep-th/0605014}{{\ttfamily hep-th/0605014}}].

\bibitem{Urrestilla:2007yw}
J.~Urrestilla and A.~Vilenkin, \emph{{Evolution of cosmic superstring networks: A Numerical simulation}}, \href{https://doi.org/10.1088/1126-6708/2008/02/037}{\emph{JHEP} {\bfseries 02} (2008) 037} [\href{https://arxiv.org/abs/0712.1146}{{\ttfamily 0712.1146}}].

\bibitem{Mandelstam:1974pi}
S.~Mandelstam, \emph{{Vortices and Quark Confinement in Nonabelian Gauge Theories}}, \href{https://doi.org/10.1016/0370-1573(76)90043-0}{\emph{Phys. Rept.} {\bfseries 23} (1976) 245}.

\bibitem{tHooft:1977nqb}
G.~'t~Hooft, \emph{{On the Phase Transition Towards Permanent Quark Confinement}}, \href{https://doi.org/10.1016/0550-3213(78)90153-0}{\emph{Nucl. Phys. B} {\bfseries 138} (1978) 1}.

\bibitem{Seiberg:1994rs}
N.~Seiberg and E.~Witten, \emph{{Electric - magnetic duality, monopole condensation, and confinement in N=2 supersymmetric Yang-Mills theory}}, \href{https://doi.org/10.1016/0550-3213(94)90124-4}{\emph{Nucl. Phys. B} {\bfseries 426} (1994) 19} [\href{https://arxiv.org/abs/hep-th/9407087}{{\ttfamily hep-th/9407087}}].

\bibitem{Seiberg:1994aj}
N.~Seiberg and E.~Witten, \emph{{Monopoles, duality and chiral symmetry breaking in N=2 supersymmetric QCD}}, \href{https://doi.org/10.1016/0550-3213(94)90214-3}{\emph{Nucl. Phys. B} {\bfseries 431} (1994) 484} [\href{https://arxiv.org/abs/hep-th/9408099}{{\ttfamily hep-th/9408099}}].

\bibitem{Donagi:1995cf}
R.~Donagi and E.~Witten, \emph{{Supersymmetric Yang-Mills theory and integrable systems}}, \href{https://doi.org/10.1016/0550-3213(95)00609-5}{\emph{Nucl. Phys. B} {\bfseries 460} (1996) 299} [\href{https://arxiv.org/abs/hep-th/9510101}{{\ttfamily hep-th/9510101}}].

\bibitem{Dorey:1999sj}
N.~Dorey, \emph{{An Elliptic superpotential for softly broken N=4 supersymmetric Yang-Mills theory}}, \href{https://doi.org/10.1088/1126-6708/1999/07/021}{\emph{JHEP} {\bfseries 07} (1999) 021} [\href{https://arxiv.org/abs/hep-th/9906011}{{\ttfamily hep-th/9906011}}].

\bibitem{Naculich:2001us}
S.~G. Naculich, H.~J. Schnitzer and N.~Wyllard, \emph{{Vacuum states of N=1* mass deformations of N=4 and N=2 conformal gauge theories and their brane interpretations}}, \href{https://doi.org/10.1016/S0550-3213(01)00291-7}{\emph{Nucl. Phys. B} {\bfseries 609} (2001) 283} [\href{https://arxiv.org/abs/hep-th/0103047}{{\ttfamily hep-th/0103047}}].

\bibitem{Vincent:1997cx}
G.~Vincent, N.~D. Antunes and M.~Hindmarsh, \emph{{Numerical simulations of string networks in the Abelian Higgs model}}, \href{https://doi.org/10.1103/PhysRevLett.80.2277}{\emph{Phys. Rev. Lett.} {\bfseries 80} (1998) 2277} [\href{https://arxiv.org/abs/hep-ph/9708427}{{\ttfamily hep-ph/9708427}}].

\bibitem{Yamaguchi:1998iv}
M.~Yamaguchi, J.~Yokoyama and M.~Kawasaki, \emph{{Numerical analysis of formation and evolution of global strings in ( 2+1)-dimensions}}, \href{https://doi.org/10.1143/PTP.100.535}{\emph{Prog. Theor. Phys.} {\bfseries 100} (1998) 535} [\href{https://arxiv.org/abs/hep-ph/9808326}{{\ttfamily hep-ph/9808326}}].

\bibitem{Yamaguchi:1999yp}
M.~Yamaguchi, \emph{{Scaling property of the global string in the radiation dominated universe}}, \href{https://doi.org/10.1103/PhysRevD.60.103511}{\emph{Phys. Rev. D} {\bfseries 60} (1999) 103511} [\href{https://arxiv.org/abs/hep-ph/9907506}{{\ttfamily hep-ph/9907506}}].

\bibitem{Press:1989yh}
W.~H. Press, B.~S. Ryden and D.~N. Spergel, \emph{{Dynamical Evolution of Domain Walls in an Expanding Universe}}, \href{https://doi.org/10.1086/168151}{\emph{Astrophys. J.} {\bfseries 347} (1989) 590}.

\bibitem{Fleury:2015aca}
L.~Fleury and G.~D. Moore, \emph{{Axion dark matter: strings and their cores}}, \href{https://doi.org/10.1088/1475-7516/2016/01/004}{\emph{JCAP} {\bfseries 01} (2016) 004} [\href{https://arxiv.org/abs/1509.00026}{{\ttfamily 1509.00026}}].

\bibitem{Hindmarsh:2017qff}
M.~Hindmarsh, J.~Lizarraga, J.~Urrestilla, D.~Daverio and M.~Kunz, \emph{{Scaling from gauge and scalar radiation in Abelian Higgs string networks}}, \href{https://doi.org/10.1103/PhysRevD.96.023525}{\emph{Phys. Rev. D} {\bfseries 96} (2017) 023525} [\href{https://arxiv.org/abs/1703.06696}{{\ttfamily 1703.06696}}].

\bibitem{Hindmarsh:2019csc}
M.~Hindmarsh, J.~Lizarraga, A.~Lopez-Eiguren and J.~Urrestilla, \emph{{Scaling Density of Axion Strings}}, \href{https://doi.org/10.1103/PhysRevLett.124.021301}{\emph{Phys. Rev. Lett.} {\bfseries 124} (2020) 021301} [\href{https://arxiv.org/abs/1908.03522}{{\ttfamily 1908.03522}}].

\bibitem{Efstratiou:2026kcb}
D.~Efstratiou, E.~A. Paraskevas and L.~Perivolaropoulos, \emph{{The friction-era VOS amplitude of a $\mathbb{Z}_2$ string network from nematic disclination data}},  \href{https://arxiv.org/abs/2608.20449}{{\ttfamily 2608.20449}}.

\bibitem{Gorghetto:2018myk}
M.~Gorghetto, E.~Hardy and G.~Villadoro, \emph{{Axions from Strings: the Attractive Solution}}, \href{https://doi.org/10.1007/JHEP07(2018)151}{\emph{JHEP} {\bfseries 07} (2018) 151} [\href{https://arxiv.org/abs/1806.04677}{{\ttfamily 1806.04677}}].

\bibitem{Ringeval:2017eww}
C.~Ringeval and T.~Suyama, \emph{{Stochastic gravitational waves from cosmic string loops in scaling}}, \href{https://doi.org/10.1088/1475-7516/2017/12/027}{\emph{JCAP} {\bfseries 12} (2017) 027} [\href{https://arxiv.org/abs/1709.03845}{{\ttfamily 1709.03845}}].

\bibitem{Chang:2021afa}
C.-F. Chang and Y.~Cui, \emph{{Gravitational waves from global cosmic strings and cosmic archaeology}}, \href{https://doi.org/10.1007/JHEP03(2022)114}{\emph{JHEP} {\bfseries 03} (2022) 114} [\href{https://arxiv.org/abs/2106.09746}{{\ttfamily 2106.09746}}].

\bibitem{Vachaspati:1984dz}
T.~Vachaspati and A.~Vilenkin, \emph{{Formation and Evolution of Cosmic Strings}}, \href{https://doi.org/10.1103/PhysRevD.30.2036}{\emph{Phys. Rev. D} {\bfseries 30} (1984) 2036}.

\bibitem{Strobl:1996yj}
K.~Strobl, \emph{{Improvements for Vachaspati-Vilenkin type algorithms for cosmic string and disclination formation}},  \href{https://arxiv.org/abs/hep-lat/9608085}{{\ttfamily hep-lat/9608085}}.

\end{thebibliography}\endgroup

\end{document}